\documentclass[lettersize,journal]{IEEEtran}
\usepackage{amsmath,amsfonts}
\usepackage{algorithmic}
\usepackage{algorithm}
\usepackage{array}
\usepackage{textcomp}
\usepackage{stfloats}
\usepackage{url}
\usepackage{verbatim}
\usepackage{graphicx}
\usepackage{subcaption}
\usepackage{tikz}
\usepackage{booktabs}
\usepackage{xcolor}
\usepackage{enumitem}
\usepackage{multirow}
\usepackage{cite}
\newcommand{\tool}{\textsc{ParDef}}

\newcommand{\bin}[1]{{\color{black} #1}} 

\begin{document}

\title{Toward a Generalized Defense Across Sparse, Continuous, and Structured Parameter Attacks}

\author{
  Bin Duan,
  Zeyu Bai,
  Guowei Yang* \\
  School of Electrical Engineering and Computer Science, The University of Queensland, Australia \\
  \{b.duan, guowei.yang\}@uq.edu.au, zeyu.bai@student.uq.edu.au
}

\markboth{Transactions on Dependable and Secure Computing}%
{Toward a Generalized Defense Across Sparse, Continuous, and Structured Parameter Attacks}

\maketitle

\begin{abstract}
Deep neural networks are increasingly deployed across heterogeneous and partially untrusted environments, where models are distributed through cloud storage, CI/CD pipelines, containerized services, and edge execution platforms. This broad deployment landscape exposes model parameters to various integrity risks. Unlike input-space adversarial attacks, parameter attacks directly tamper with the model’s internal parameters and persist across all subsequent inferences. Existing defenses either require retraining, incur significant accuracy degradation, or are limited to specific attack classes. However, in real-world deployment scenarios, the forms of parameter attacks are often unpredictable. To address this challenge, we present \tool, a generalized defense for deep neural networks against diverse types of parameter attacks.
\tool\ integrates keyed channel reparameterization, which obscures sensitive parameter directions, QC-LDPC quantization, which embeds redundancy and supports error correction, and adaptive robust inference, which stabilizes predictions under uncertainty.
Our evaluation on CIFAR-10, CIFAR-100, and Tiny-ImageNet using ResNet and VGG models\bin{, together with additional DeiT experiments on ImageNet-1K and CIFAR-100,} demonstrates that \tool\ consistently reduces attack success rates across different parameter attacks while maintaining high model performance and incurring only moderate deployment overhead. 
These results highlight that \tool\ is a practical and generalized defense for \bin{securing at-rest model parameters} in DNN deployments.

\end{abstract}


\section{Introduction}
\IEEEPARstart{D}{eep} neural networks (DNNs) have become integral to modern computing and service infrastructures, powering core tasks such as classification~\cite{krizhevsky2012imagenet}, search and ranking~\cite{huang2013learning}, recommendation~\cite{cheng2016wide}, and large-scale personalization~\cite{covington2016deep}. 
Billions of inferences are executed daily across cloud platforms, edge devices, distributed model-serving pipelines, and containerized microservices~\cite{crankshaw2017clipperserving}, reflecting both the scalability of DNN-based applications and the expanding security surface of deployed models~\cite{yu2022modelinv}.

In controlled server-side settings, deployed models typically operate within a narrow and well-monitored trust boundary~\cite{babar2023trusted}. 
However, modern ML deployment pipelines increasingly distribute models across heterogeneous environments—cloud storage services, CI/CD delivery channels, container images, edge endpoints, and shared model repositories—substantially enlarging the attack surface for parameter manipulation~\cite{asmus2023supplychain}. 
Compromised or maliciously modified models have already been observed in public repositories, including persistent backdoored parameters that survive fine-tuning~\cite{kurita2020backdoor}, as well as parameter-level manipulations designed to leak private information or induce targeted misbehavior~\cite{gan2024navigating}. 
Audits of large-scale model hubs further indicate that unsafe serialization and model packaging practices can expose critical integrity vulnerabilities~\cite{wen2024pickle}. 
In operational ML pipelines, untrusted artifacts and automated model-retrieval processes have been identified as practical injection vectors for tampered parameters~\cite{wasay2022morello}, highlighting the need for principled mechanisms that ensure model integrity throughout deployment.

Parameter attacks exploit this expanded attack surface by directly corrupting the stored model parameters to induce targeted failures~\cite{hou2024ibd}. These attacks {can be classified into three distinct types}: sparse bit-flip attacks, which target critical bits in floating-point representations~\cite{rakin2019bitflip}, continuous bounded noise injection across weight tensors~\cite{liu2019faultsneak,zhang2020advbit}, and structured manipulations, which exploit architectural regularities in channels or layers~\cite{yao2020deephammer,zhao2021flip}. Unlike input-space attacks~\cite{fang20253sat}, which can sometimes be {mitigated} through preprocessing, parameter attacks operate stealthily within the model's computational core and persist across all subsequent inferences~\cite{he2020towards}.

The impact of parameter attacks is amplified in modern deployment settings, where models are frequently distributed across heterogeneous and partially untrusted execution environments~\cite{ali2024rise}. 
Once a compromised model is propagated through deployment pipelines or distributed endpoints, the tampered parameters persist and can rapidly affect a large number of downstream applications~\cite{mitropoulos2017defending}. 
Existing defenses, such as aggressive quantization (BIN~\cite{rakin2019bit}, RA-BNN~\cite{he2020defending}), incur substantial accuracy degradation and require full retraining, which is often impractical in real deployment workflows that demand efficiency and minimal disruption.
Recent specialized defenses like Aegis~\cite{wang2023aegis} provide strong protection against bit-flip attacks through multi-exit architectures, achieving notable reductions in attack success rates. 
However, such approaches remain narrowly tailored to sparse parameter perturbations and offer limited robustness against continuous or structured parameter attacks, leaving significant portions of the threat landscape unaddressed.

In this paper, we propose \tool, a generalized defense against diverse parameter attacks, providing robustness to sparse, continuous, and structured attacks.
\tool\ incorporates three synergistic mechanisms: (1) keyed channel reparameterization~(KCR), which reparameterizes parameter channels using secret keys to obscure gradient-based attack directions, (2) quasi-cyclic low-density parity-check (QC-LDPC)~\cite{li2006efficient} coded quantization, which embeds error-correcting redundancy while reducing model size, and (3) adaptive robust inference~(ARI), which dynamically allocates computational resources based on prediction uncertainty.
Together, these mechanisms enable \tool\ to achieve broad robustness without retraining, maintain functional equivalence with the original model, remain compatible with existing inference frameworks, and reduce storage overhead while preserving inference efficiency.

We evaluate \tool\ on CIFAR-10~\cite{krizhevsky2009learning}, CIFAR-100, and Tiny-ImageNet~\cite{deng2009imagenet}. 
{\tool\ consistently defends against the three types of parameter attacks, achieving lower attack success rates than existing defenses while maintaining high model performance and incurring minimal deployment overhead. \bin{We further conduct an additional DeiT-based evaluation on ImageNet-1K and CIFAR-100 to examine Transformer compatibility and overhead scaling.} These results position \tool\ as a practical and generalized defense for DNNs against diverse parameter attacks.}
To support reproducibility, we release artifacts~\cite{Our}.

\section{Background}

\subsection{Parameter Attacks}

Parameter attacks are an emerging class of adversarial attacks that manipulate the internal parameters of DNNs, altering model functionality without changing any inputs~\cite{rakin2019bit,yao2020deephammer}.  
Unlike input-based attacks, which perturb data and can often be mitigated through training or preprocessing, parameter attacks strike at the model’s core, making them stealthier and more destructive: they evade input-level defenses and persist across future inferences once parameters are compromised. Such attacks are particularly critical in modern deployment pipelines, where adversaries may access model parameters stored in shared repositories, propagated through CI/CD workflows, or distributed across edge and containerized execution environments~\cite{sood2025malicious, veria2024threat}.
From a perturbation perspective, we group parameter attacks into three categories:  
(1) {Sparse attacks} ($L_0$-bounded), modifying only a few parameters;  
(2) {Continuous attacks} ($L_2$/$L_\infty$-bounded), distributing bounded noise across many parameters;  
(3) {Structured attacks}, introducing correlated changes across groups, channels, or layers.  
Here, $L_0$ measures the number of altered parameters, $L_2$ their Euclidean magnitude, and $L_\infty$ the maximum per-parameter change. These categories capture most parameter attacks~\cite{shuvo2023comprehensive}.

\noindent \textbf{\textit{Sparse parameter attacks ($L_0$-bounded).}}
These attacks modify only a small subset of weights, keeping the rest intact to maximize stealth and minimize functional disturbance until the intended trigger. Representative examples include bit-flip attacks~(BFAs) such as targeted bit-flip attacks and ProFlip~\cite{chen2021proflip}, which exploit the binary encoding of parameters, flipping exponent or high-order mantissa bits can induce large numerical changes. Targeted bit-flip attack ranks bits by gradient-based loss impact, while ProFlip uses probabilistic search to minimize the number of flips. 

\noindent \textbf{\textit{Continuous parameter attacks ($L_2$/$L_\infty$-bounded).}}  
Here, adversaries apply small but dense additive noise across many parameters under norm constraints, such as $\|\delta W\|_2 \le \epsilon$ or $\|\delta W\|_\infty \le \epsilon$.  
The Adversarial Parameter Attack (APA)~\cite{rakin2019bit} perturbs loss-sensitive parameters using gradient guidance, causing severe accuracy drops while preserving global structure.  
This category also includes random or adversarial noise injection mimicking quantization errors or hardware faults.

\noindent \textbf{\textit{Structured parameter attacks.}}
These attacks target entire groups, filters, or layers, exploiting network hierarchy to maximize impact.  
The Parameter Adaptive Adversarial Attack (P3A)~\cite{yao2020deephammer} exemplifies this strategy by adaptively tuning attacks toward vulnerable substructures, yielding large-scale degradation while evading naive retraining defenses.

These attacks can cause catastrophic accuracy drops~\cite{rakin2019bitflip}, targeted misclassification, or persistent backdoor behavior~\cite{yao2020deephammer}, all while leaving the model’s input-output interface unchanged. 
This breadth of attack highlights the need for generalized defenses capable of addressing all attack types without relying on attack-specific assumptions, especially given the wide range of deployment environments where models may be exposed to adversarial manipulation.

\subsection{Defenses Against Parameter Attacks}

Parameter-attack defenses generally fall into two categories: integrity verification and model enhancement. Integrity verification methods, such as error-correcting schemes~\cite{liu2020fault,hong2020terminal}, introduce redundancy into parameter storage to detect or repair corrupted weights. While effective in hardware-aware contexts, these approaches require specialized support, incur computational and memory overhead, and remain reactive rather than preventive.
In contrast, model enhancement methods directly modify the model to improve robustness against parameter attacks. BIN~\cite{rakin2019bit} and RA-BNN~\cite{he2020defending} restrict parameters to binary or quantized values, thereby limiting the possible impact of bit-level changes. However, such constraints often degrade clean accuracy and require complete retraining, which reduces practicality for large-scale or legacy models. More recently, {Aegis}~\cite{wang2023aegis} defenses targeted bit-flip attacks via a dynamic-exit architecture and robustness-oriented training, but introduces notable architectural/training overhead and is validated on BFA-style attacks. BITSHIELD~\cite{chen2025bitshield} focuses on compiled DNN executables, detecting BFAs on data and code sections.
Importantly, the demonstrated robustness of these defenses is largely confined to targeted bit-flip scenarios; their generality against other parameter attacks remains uncertain.

\begin{figure*}[t!]
\centering
\includegraphics[width=0.75\textwidth]{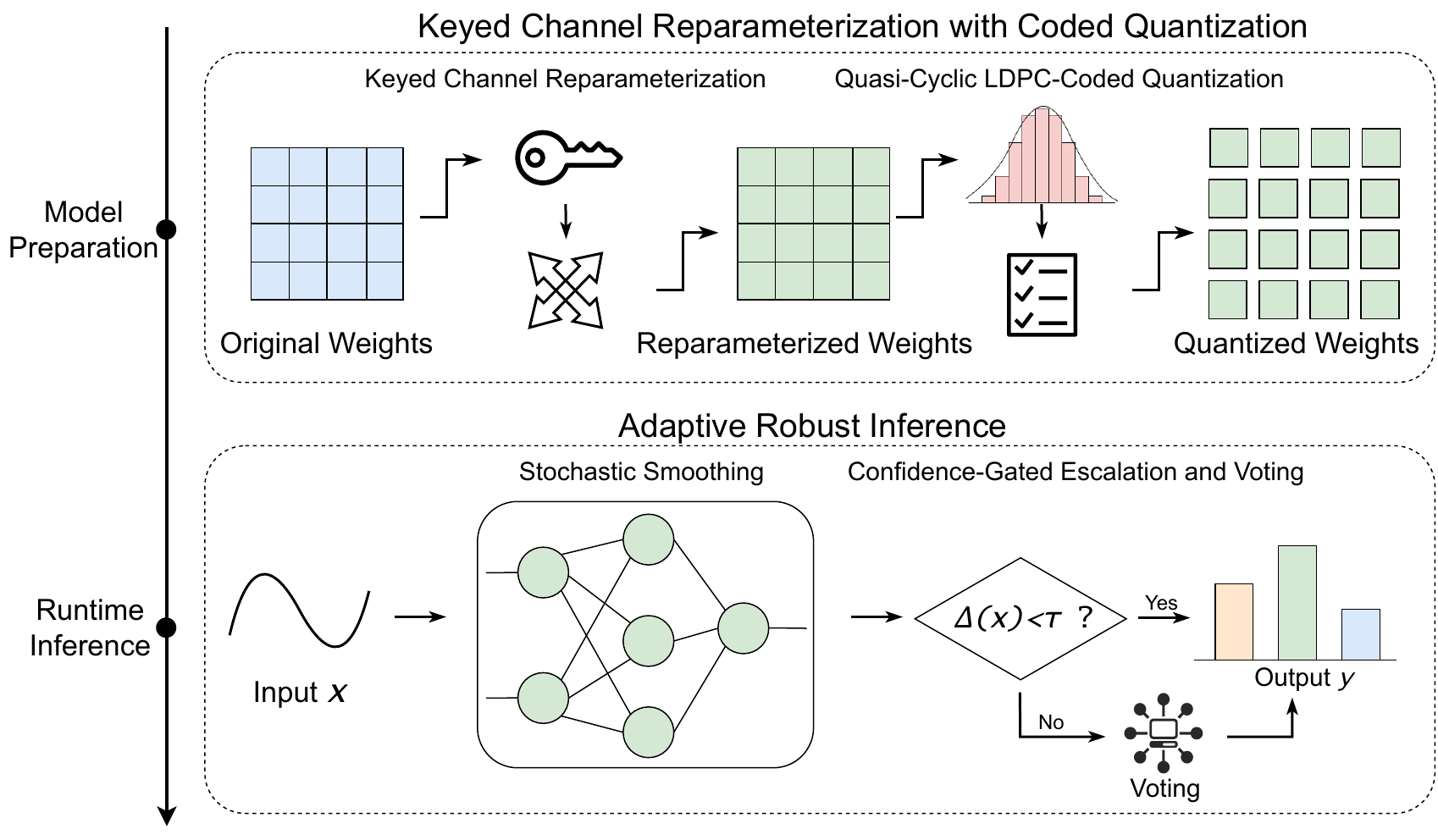}
\vspace{-2mm}
\caption{Overview of \tool.}
\vspace{-4mm}
\label{fig:overview}
\end{figure*}

\section{Threat Model and Defense Requirements}

\subsection{Threat Model}
\label{sec:threat-model}
We consider a parameter-attack adversarial scenario in which the adversary can directly modify the stored parameters of a deployed DNN model before inference. This setting is motivated by a growing body of evidence~\cite{rakin2019bit,yao2020deephammer,shuvo2023comprehensive} showing that model files, cached artifacts, and edge-deployed DNNs are vulnerable to weight-level corruption through supply-chain manipulation, unsafe model distribution, or compromised deployment artifacts. \bin{Runtime vectors such as row-hammer style faults or co-located malicious processes further illustrate the practical severity of parameter-level corruption, but our primary focus is at-rest parameter tampering before model loading, as clarified below.} Even a small number of parameter changes can cause catastrophic model failures, making parameter attacks a realistic and severe threat.

\noindent\textbf{Attack surface focus.}
We focus on {at-rest} (pre-load) parameter tampering, where an adversary modifies the serialized model checkpoint before the model is loaded for inference. We consider {post-load} in-memory parameter tampering (after model initialization) as out of scope, because it bypasses load-time verification/decoding and requires orthogonal runtime integrity mechanisms. 

\noindent\textbf{Trusted computing base and key protection.}
\bin{Following established practice in machine learning systems security that builds on trusted execution environment (TEE) primitives~\cite{tramer2019slalom, mo2020darknetz}, \tool\ adopts a tiered trust model and achieves its strongest guarantees when a minimal TEE (e.g., Intel SGX~\cite{mckeen2013intel} or ARM TrustZone~\cite{cerdeira2020sok}) is available on the inference platform (see Section~\ref{sec:deployment-tiers}).} \bin{In Tier~1,} \tool's preparation-time secrets, including the KCR permutations and scaling factors, are generated and stored inside the TEE and never exposed outside of it. The adversary can fully access and modify only the transformed model parameters stored outside the TEE, but cannot read or tamper with the secrets or code residing inside the trusted environment. This prevents direct key extraction~\cite{yeom2018privacy,russinovich2021confidential} while still allowing strong parameter-space attacks on the deployed model~\cite{carlini2020cryptanalytic,shamir2023polynomial}. We assume the TEE provides standard isolation guarantees and do not consider side-channel leakage or physical compromise of the trusted environment, which are orthogonal to the parameter-attack threat studied in this work.

\noindent\textbf{Adversarial capabilities.}
Following prior work on strong parameter-space adversaries~\cite{chen2021proflip,yao2020deephammer,wang2023aegis}, we assume a {white-box, defense-aware} adversary who has full access to the final deployed model, including:
(i)~its architecture,
(ii)~all model parameters after applying our defense mechanism, and
(iii)~the complete inference-time code path, excluding the preparation pipeline executed inside the TEE.
The adversary does not observe any internal artifacts from the preparation pipeline, but can freely inspect and analyze the final transformed model exactly as deployed.

The adversary may modify model parameters within a fixed attack budget and may adopt any manipulation strategy considered in prior Section~\ref{sec:threat-model}, including:
sparse ($L_0$-bounded) perturbations,
continuous ($L_2/L_\infty$-bounded) perturbations,
and structured channel- or layer-aligned perturbations.  
The adversary is allowed to compute gradients or sensitivity estimates on the deployed model, to perform quantization-aware or structure-aware optimization, and to use Expectation-over-Transformation (EOT) techniques to adapt to stochastic inference.  
However, the adversary cannot interfere with the runtime platform, nor can they access the original, untransformed model used before deployment.

\noindent\textbf{Adversarial goals.}  
Following prior work on parameter-space attacks, we consider adversaries that manipulate model parameters to induce prediction failures, as we mentioned in Section~\ref{sec:threat-model}.  
Depending on the attack strategy, the adversary may pursue one of the following goals:

\begin{itemize}[leftmargin=*]

\item \textbf{Sparse corruption.}  
The adversary modifies a very small number of highly sensitive parameters, typically via selective bit flips, so as to maximally disrupt model predictions with minimal perturbation.  
Such sparse attacks are known to cause significant functional shifts even when only a few parameters are altered~\cite{wang2023aegis, rakin2019bit}.

\item \textbf{Continuous perturbation.}  
The adversary introduces small but widespread perturbations across many parameters, constrained by $L_2$ or $L_\infty$ norms, with the goal of gradually deteriorating the model’s stability and decision boundaries.  
Such dense perturbations emulate continuous parameter drift or fine-grained adversarial noise~\cite{rakin2019bit}.

\item \textbf{Structured disruption.}  
The adversary perturbs parameters in a coordinated manner across filters, channels, or layers, aiming to corrupt intermediate representations or feature pathways.  
By targeting structurally related parameter groups, these attacks can induce large-scale degradation while preserving stealth~\cite{yao2020deephammer}.

\end{itemize}

These adversarial capabilities and goals encompass the primary forms of realistically executable parameter attacks demonstrated in prior section and define the threat surface addressed by \tool.

\subsection{Defense Requirements}

Our goal is to design a practical defense that provides robust protection across representative families of parameter attacks, including adaptive and defense-aware variants.
Rather than attempting to make parameter attacks theoretically impossible, the defense should increase the adversary’s required corruption budget and reduce the feasibility of successful parameter manipulation in real-world deployment settings.
To achieve this, the defense should satisfy the following requirements:

\begin{itemize}[leftmargin=*]

\item \textbf{Non-intrusive.}
This defense should be applicable to fully trained DNN models without requiring any retraining, enabling that the mechanism can be integrated into existing deployments.

\item \textbf{Utility-preserving.}
The defense should preserve the model’s predictive performance on clean inputs and avoid degrading accuracy or decision stability.  
A practical defense should not highly compromise the original functionality or reliability of the model.

\item \textbf{Low-overhead.}
The defense should introduce minimal computational, storage, and inference-time overhead, ensuring that the protected model remains efficient and deployable in real-world systems.  
Both model size and latency should remain close to those of the original model.

\item \textbf{Robust under adaptive attacks.}
The defense should remain effective even when adversaries are defense-aware and perform gradient-based, quantization-aware, or EOT-based optimization on the {deployed} model.  
This requirement ensures that robustness does not rely on gradient obfuscation or unrealistic attacker limitations.

\item \textbf{Broad attack coverage.}  
The defense should offer robustness across diverse parameter-attack strategies.  
Real-world parameter attacks may be sparse, continuous, or structured.  
A practical defense should withstand these three representative manipulation forms, in order to avoid leaving openings for adaptive adversaries who simply switch attack patterns.

\end{itemize}

\section{Method}

{

This section presents \tool, a defense pipeline designed to provide broad protection across three representative families of parameter attacks, sparse ($L_0$), continuous ($L_2/L_\infty$), and structured, without relying on additional loss terms or retraining.
Figure~\ref{fig:overview} shows the overview of \tool, which integrates three complementary modules spanning both model preparation and runtime inference.
(1)~\textit{Keyed Channel Reparameterization} (KCR) obfuscates and diffuses sensitive directions in parameter space while preserving model functionality~(Sec.~\ref{sec:KCR}).
(2)~\textit{QC-LDPC Coded Quantization} constrains model weights to redundancy-protected codewords, enabling correction or detection of both sparse bit flips and small continuous drifts~(Sec.~\ref{sec:QC-LDPC}).
(3)~\textit{Adaptive Robust Inference} (ARI) provides runtime stability by combining stochastic smoothing with confidence-gated redundancy~(Sec.~\ref{sec:ARI}).
Together, these mechanisms form a preparation--inference defense pipeline that increases the attacker’s required corruption budget across the three attack families considered.
}

\subsection{Keyed Channel Reparameterization}
\label{sec:KCR}

Keyed Channel Reparameterization~(KCR) reindexes and rescales channels in a key-conditioned way to misalign adversarial gradients and diffuse localized perturbations, while exactly preserving the function of the model.
Intuitively, KCR hinders an attacker’s ability to target the most ``sensitive'' neurons or channels, since the deployed basis is key-randomized and instantiated by secrets kept inside the TEE (the adversary knows the mechanism but not the key).

Specifically, for each convolutional or linear layer $l$ with a weight tensor or matrix
$W_l \in \mathbb{R}^{C_{\text{out}}\times C_{\text{in}}\times k_h \times k_w}$ (linear layer: $k_h{=}k_w{=}1$) and bias $b_l\in\mathbb{R}^{C_{\text{out}}}$, we sample two key-driven, invertible channel transforms in restricted form:
\begin{equation}
P_l=\Pi^{(\mathrm{out})}_l D^{(\mathrm{out})}_l,\quad
Q_l=\Pi^{(\mathrm{in})}_l D^{(\mathrm{in})}_l,\quad D^{(\cdot)}_l \succ 0,
\end{equation}
where $\Pi^{(\cdot)}_l$ are permutation matrices and $D^{(\cdot)}_l$ are positive diagonal scalings. Positivity ensures reversibility and avoids sign flips around activations.
Using $(P_l,Q_l)$, the transformed parameters are
\begin{equation}
\label{eq:kcr-matrix-compact}
\begin{aligned}
\widetilde{W}_l = P_l\, W_l\, Q_l^{-1},
\widetilde{b}_l = P_l\, b_l .
\end{aligned}
\end{equation}

Equivalently, unfolding $P_l$ and $Q_l$ into $\Pi,D$, when $k_hk_w>1$ we can view $W_l$ in matrix form and recover
\begin{equation}
\label{eq:kcr-matrix}
\widetilde{W}_l 
= D^{(\mathrm{out})}_l\,\Pi^{(\mathrm{out})}_l\; W_l\; \Pi^{(\mathrm{in})\top}_l\,[D^{(\mathrm{in})}_l]^{-1},
\end{equation}
and in indexed tensor form
\begin{equation}
\label{eq:kcr-weight}
\begin{aligned}
&\widetilde{W}_l(o,i,k_h,k_w) \\
&= D^{(\mathrm{out})}_l(o)\;
  W_l\!\big(\Pi^{(\mathrm{out})}_l(o),\,\Pi^{(\mathrm{in})}_l(i),\,k_h,\,k_w\big)\;
  [D^{(\mathrm{in})}_l(i)]^{-1},
\end{aligned}
\end{equation}
with bias
\begin{equation}
\label{eq:kcr-bias}
\widetilde{b}_l \;=\; D^{(\mathrm{out})}_l\,\Pi^{(\mathrm{out})}_l\, b_l .
\end{equation}

To keep tensors consistent with the reparameterized weights, the input and output features of layer $l$ are mapped as
\begin{equation}
\label{eq:kcr-feature}
\begin{aligned}
\widetilde{x}_l = Q_l\, x_l,
\widetilde{y}_l = P_l\, y_l .
\end{aligned}
\end{equation}
where $x_l$ is the input to $W_l$ and $y_l$ is the pre-normalization output of $W_l$ in the original network. This mapping will be implemented implicitly by updating adjacent normalizations and by propagating permutations across edges of the computation graph.

Let the original per-channel affine normalizations be
\[
\begin{aligned}
N_{\mathrm{in}}(x)  = A_{\mathrm{in}} x + c_{\mathrm{in}},
N_{\mathrm{out}}(z) = A_{\mathrm{out}} z + c_{\mathrm{out}}.
\end{aligned}
\]
with $A_{\mathrm{in}},A_{\mathrm{out}}$ diagonal and $c_{\mathrm{in}},c_{\mathrm{out}}$ vectors. Under KCR, choose updated normalizations:
\begin{equation}
\label{eq:kcr-affines-compact}
\begin{aligned}
&\widetilde{A}_{\mathrm{in}} = Q_l\, A_{\mathrm{in}},\qquad
  \widetilde{c}_{\mathrm{in}} = Q_l\, c_{\mathrm{in}},\\
&\widetilde{A}_{\mathrm{out}} = P_l\, A_{\mathrm{out}}\, P_l^{-1},\qquad
  \widetilde{c}_{\mathrm{out}} = P_l\, c_{\mathrm{out}}.
\end{aligned}
\end{equation}
Thus
\[
\widetilde{N}_{\mathrm{in}}(x)=\widetilde{A}_{\mathrm{in}}x+\widetilde{c}_{\mathrm{in}}
= Q_l\,(A_{\mathrm{in}}x+c_{\mathrm{in}})=Q_l\,N_{\mathrm{in}}(x),
\]
\[
\begin{aligned}
\widetilde{N}_{\mathrm{out}}(z)
= \widetilde{A}_{\mathrm{out}}\, z + \widetilde{c}_{\mathrm{out}} 
&= P_l\!\left(A_{\mathrm{out}}(P_l^{-1}z)+c_{\mathrm{out}}\right) \\[2pt]
&= P_l\, N_{\mathrm{out}}(P_l^{-1}z).
\end{aligned}
\]

\noindent\textbf{Layer-wise Equivalence Proof.}
Consider the inference-time block
\[
x \xrightarrow{N_{\mathrm{in}}}\; z = W_l x + b_l \xrightarrow{N_{\mathrm{out}}}\; \mathrm{Act}(\cdot).
\]
Using (\ref{eq:kcr-feature})–(\ref{eq:kcr-affines-compact}) and $\widetilde{W}_l, \widetilde{b}_l$ from (\ref{eq:kcr-matrix-compact}), we have
\[
\begin{aligned}
\widetilde{W}_l\, \widetilde{N}_{\mathrm{in}}(x) + \widetilde{b}_l
&= (P_l W_l Q_l^{-1})\, (Q_l N_{\mathrm{in}}(x)) + P_l b_l\\[2pt]
&= P_l \big(W_l N_{\mathrm{in}}(x) + b_l\big).
\end{aligned}
\]
Applying the updated output normalization,
\[
\widetilde{N}_{\mathrm{out}}\!\big(\widetilde{W}_l\, \widetilde{N}_{\mathrm{in}}(x) + \widetilde{b}_l\big)
= P_l \Big( N_{\mathrm{out}}\!\big(W_l\, N_{\mathrm{in}}(x) + b_l\big)\Big).
\]
Finally, since $\mathrm{Act}$ is channel-wise, it is permutation-equivariant:
$\mathrm{Act}(\Pi^{(\mathrm{out})}_l u)=\Pi^{(\mathrm{out})}_l\,\mathrm{Act}(u)$.
Under the activation constraint stated below (i.e., diagonal scalings are set to identity
across non-homogeneous activations), we obtain
\begin{equation}
\label{eq:kcr-layer-lemma}
\mathrm{Act}\!\circ\widetilde{N}_{\mathrm{out}}\!\big(\widetilde{W}_l\, \widetilde{N}_{\mathrm{in}}(x) + \widetilde{b}_l\big)
= P_l \,\Big(\mathrm{Act}\!\circ N_{\mathrm{out}}\Big)\!\big(W_l\, N_{\mathrm{in}}(x) + b_l\big).
\end{equation}
That is, the transformed block equals the original block up to a channel permutation given by the permutation part of $P_l$.

\noindent\textbf{Network-wise Strict Equivalence.}
Let the network be a feed-forward graph with possible residual or parallel branches. Assign KCR transforms $\{(P_l,Q_l)\}$ subject to:

\noindent (i). Edge consistency. For any edge from layer $l$ to $l{+}1$, set the consumer’s input transform equal to the producer’s output transform, i.e., $Q_{l+1}=P_l$.

\noindent (ii). Branch alignment. For residual, add, or concat nodes, apply the same permutation to all aligned branches before merging.

\noindent (iii). Endpoints. Choose $Q_{\text{first}}=I$ and $P_{\text{last}}=I$ (or append a fixed inverse permutation at the network tail).

Then all intermediate permutations cancel along edges, merged tensors remain channel-aligned, and the external input and output order is preserved; hence, for all inputs $x$,
\begin{equation}
\label{eq:kcr-network-prop}
\widetilde{f}(x)=f(x).
\end{equation}

To extend the equivalence argument to practical architectures, we specify how the reparameterization applies to common structures in modern DNNs.

\noindent \textbf{Normalization.}
For a normalization layer {before} $W_l$, compose it with $Q_l$, permuting by $\Pi^{(\mathrm{in})}_l$ and adjust its per-channel affine parameters as in (\ref{eq:kcr-affines-compact}).  
For a normalization layer {after} $W_l$, compose it with $P_l$ and its inverse, permuting by $\Pi^{(\mathrm{out})}_l$ and adjust its per-channel affine parameters so the overall affine becomes $P_l(\cdot)P_l^{-1}$.

\noindent \textbf{Activations.}
Elementwise activations (ReLU/GELU/SiLU) commute with channel permutations. We set diagonal scalings to identity across non-homogeneous activations when strict commutation with $D^{(\cdot)}_l$ is not guaranteed, and rely on permutations for those boundaries.

\noindent \textbf{Residual connections.}
If the permutation part of $P_l$, $\Pi^{(\mathrm{out})}_l$, permutes the main branch, the skip branch must apply the same permutation to maintain channelwise alignment; for a $1{\times}1$ skip,
\(
\widetilde{W}^{\mathrm{skip}}(o,i,1,1) \;=\; W^{\mathrm{skip}}(\Pi^{(\mathrm{out})}_l(o), i,1,1).
\)

\noindent \textbf{Group or depthwise convolutions.}
Permutations must respect group structure: $\Pi^{(\mathrm{in})}_l$ (and $\Pi^{(\mathrm{out})}_l$ when applicable) permute {within} each group. For depthwise conv ($C_{\mathrm{in}}\!=\!C_{\mathrm{out}}$, group$=C_{\mathrm{in}}$), tie the permutation parts so that $\Pi^{(\mathrm{out})}_l=\Pi^{(\mathrm{in})}_l$, preserving the input–output channel correspondence exactly.

Equations (\ref{eq:kcr-matrix-compact})–(\ref{eq:kcr-feature}) define parameter and feature transforms; (\ref{eq:kcr-affines-compact}) absorbs them into adjacent normalizations; the layer identity (\ref{eq:kcr-layer-lemma}) shows each block is preserved up to a permutation; the network conditions above cancel permutations globally, yielding (\ref{eq:kcr-network-prop}) and thus strict functional equivalence with zero inference overhead.

By randomizing the channel basis while maintaining strict functional equivalence, KCR disperses sparse perturbations, disrupts structured channel-wise attacks, and misaligns gradient-based optimization, increasing the difficulty of targeted parameter corruption.

\subsection{Quasi-cyclic Low-density Parity-check Coded Quantization}  
\label{sec:QC-LDPC}
While KCR diffuses and obfuscates sensitive directions, it does not provide robustness against bit-level corruption or small continuous drifts. 
To address these residual vulnerabilities, we adopt a QC-LDPC quantization scheme~\cite{li2006efficient}. 
In this scheme, the parity-check matrix is composed of repeated circulant submatrices, enabling efficient encoding and decoding while maintaining strong error-correction capability. 
By constraining weights to valid codewords, QC-LDPC corrects perturbations within the code's error-correction capability and reliably detects larger deviations, while block-wise decoding repairs sparse bit flips and mitigates small continuous drifts.

Prior to deployment, we uniformly quantized to \(b\)-bit precision using affine quantization within a per-tensor range, yielding quantization indices
\begin{equation}
q(\theta_i; b) \in \{0,1,\ldots,2^b-1\},
\end{equation}
where \(\theta_i\) denotes a weight and \(q(\theta_i; b)\) is its \(b\)-bit index, and the affine quantization is defined as
\begin{equation}
q(\theta_i; b) = \text{round}\!\left(\frac{\theta_i - \theta_{\min}}{\Delta}\right), 
\Delta = \frac{\theta_{\max} - \theta_{\min}}{2^b - 1}.
\end{equation}
where \(\theta_{\min}\) and \(\theta_{\max}\) denote the minimum and maximum values of the tensor, and \(\Delta\) is the quantization step size.

Quantized indices are grouped into fixed-size blocks ($N_b$ indices per block) and encoded via a QC-LDPC code, ensuring all stored bit sequences are valid codewords.  
Bit errors that fall within the code's correction capability are recovered~\cite{kee2024review}; larger deviations are detected at load time and trigger a fail-safe (aborting model loading), preventing silent deployment with corrupted weights. 
The quasi-cyclic structure keeps encoding and decoding efficient~\cite{chen2010memory}. All LDPC decoding occurs once at model loading time; inference uses the decoded weights without any per-query overhead.

This integration of discrete quantization and QC-LDPC block coding produces redundancy-protected parameters: correctable bit-level errors are repaired, small continuous perturbations are mapped back to a valid codeword within the correction radius, and larger deviations are localized rather than silently propagating through the network. When combined with KCR, remaining perturbations are further dispersed, making targeted parameter manipulations harder.

\subsection{Adaptive Robust Inference}
\label{sec:ARI}

ARI is an inference-time module that handles residual perturbations left after KCR and QC-LDPC by stabilizing predictions through stochastic smoothing and uncertainty-gated redundancy, without retraining or persistently altering the stored parameters.

\subsubsection{Stochastic Smoothing}  

For each incoming input $x$, we introduce lightweight, ephemeral randomized perturbations during inference by injecting small Gaussian noise $\xi \sim \mathcal{N}(0,\sigma_w^2 I)$ (with $\sigma_w$ chosen to avoid accuracy degradation). This yields perturbed parameters $W' = W + \xi$, and the model performs $M$ stochastic forward passes under independently sampled $\xi_1, \dots, \xi_M$. The resulting logits are averaged to produce a smoothed prediction:
\[
\bar{z} = \frac{1}{M}\sum_{m=1}^{M} z^{(m)}.
\]

By marginalizing over a distribution of perturbed models, ARI smooths decision boundaries in parameter space, reducing sensitivity to sparse ($L_0$-bounded) and continuous ($L_2/L_\infty$-bounded) parameter distortions. Because the adversary is allowed to apply expectation over transformation~(EOT), smoothing does not rely on gradient masking; robustness arises from variance reduction and boundary stabilization, not from obscuring gradients.

\subsubsection{Confidence-Gated Escalation and Voting}  

This mechanism allocates computation adaptively by escalating only those inputs that exhibit high decision uncertainty, thereby providing robustness without incurring unnecessary overhead on benign inputs. 

Given an input $x$, the model first performs $M_s$ stochastic forward passes. In each pass, the model parameters are perturbed by adding a small Gaussian noise $\xi \sim \mathcal{N}(0,\sigma_w^2 I)$, producing logits $\{z^{(1)}, z^{(2)}, \dots, z^{(M_s)}\}$. These logits are averaged to obtain $\bar{z}$, which is converted to probabilities $p = \mathrm{softmax}(\bar{z})$.
We define the {confidence margin} as
\begin{equation}
\Delta(x) = p_{(1)} - p_{(2)},
\end{equation}
where $p_{(1)}$ and $p_{(2)}$ are the probabilities of the most and second-most likely classes. A small $\Delta(x)$ indicates that the prediction is close to the decision boundary, where even minor parameter attacks could cause label flips. An input is flagged as {high-risk} if $\Delta(x)<\tau$, where $\tau$ is selected once on a clean validation set as a quantile of the margin distribution and then fixed for all evaluations, so that escalation is rare for benign parameters yet sensitive to parameter drift.

For high-risk inputs, we switch to a slow path with $M_\ell \gg M_s$ higher-precision stochastic passes. 
Each pass draws an independent $\xi^{(i)}$ and produces logits $z^{(i)}$ and a predicted class:
\begin{equation}
\hat y^{(i)}=\arg\max_k z^{(i)}_k. 
\end{equation}

We then form per-class vote counts 
\begin{equation}
v_k=\sum_{i=1}^{M_\ell}\mathbf{1}\{\hat y^{(i)}=k\}
\end{equation}
and return
\begin{equation}
\hat y(x)=\arg\max_{k} v_k,
\end{equation}
and breaking ties using the average softmax. 

This redundancy reduces prediction variance and suppresses transient or localized parameter corruptions, while stochastic diversity prevents correlated structured attacks from consistently biasing the majority vote. Because ARI’s stochasticity is fully visible to the adversary (who may apply EOT), robustness results from redundancy and variance reduction rather than from hiding gradients.




\section{Evaluation}
\subsection{Research Questions}

Our evaluation is guided by four research questions:

\begin{itemize}
  \item[\textbf{RQ1:}] Does \tool\ preserve the clean predictive performance of the original model?
  \item[\textbf{RQ2:}] How effective is \tool\ in mitigating diverse parameter attacks?
  \item[\textbf{RQ3:}] What deployment overhead does \tool\ introduce in terms of model size and inference latency?
  \item[\textbf{RQ4:}] How do the individual components of \tool\ contribute to its overall robustness?
\end{itemize}

{
For \textbf{RQ1}, we examine how \tool\ affects clean accuracy to evaluate its utility preservation.
For \textbf{RQ2}, we measure robustness under three representative attack types, sparse ($L_0$), continuous ($L_2/L_\infty$), and structured perturbations, together with their adaptive variants \bin{and jointly optimized adaptive variants}, to assess the breadth of attack coverage.
For \textbf{RQ3}, we analyze deployment costs by measuring changes in model size and runtime latency \bin{under both typical and worst-case slow-path settings}.
For \textbf{RQ4}, we conduct ablation studies by removing KCR, LDPC-coded quantization, and ARI individually \bin{and in combination}, to quantify the contribution of each component \bin{and their cross-module complementarity} to overall robustness.
}

\subsection{Experimental Setup}

\noindent\textbf{Datasets and models.} 
We followed~\cite{wang2023aegis, rakin2019bit, he2020defending} and conducted our experiments on three widely used datasets with two DNN models: VGG16 and ResNet32.

\noindent \textbf{CIFAR-10 \& CIFAR-100}~\cite{krizhevsky2009learning}: Both datasets consist of 50,000 training images and 10,000 test images of size $32 \times 32 \times 3$. The difference lies in the number of categories: CIFAR-10 includes 10 classes, while CIFAR-100 contains 100 classes.

\noindent \textbf{Tiny-ImageNet}~\cite{deng2009imagenet}: This dataset is a simplified version of ImageNet consisting of color images with a size of $64 \times 64 \times 3$ belonging to 200 classes. 

We separate the training and testing data without any overlap. 
For {CIFAR-10 and CIFAR-100}: we select 50,000/10,000 images for training/testing. 
For {Tiny-ImageNet}: we select 100,000/10,000 images for training/testing.

\noindent\textbf{Baselines.}
We compare \tool\ against four baselines that represent complementary defense paradigms:
\textbf{Base}, which is the unprotected model;
\textbf{Aegis}~\cite{wang2023aegis}, which employs a dynamic multi-exit architecture with robustness-oriented training against targeted bit-flip attacks;
\textbf{BIN}~\cite{rakin2019bit}, which constrains parameters to binary values to limit the impact of bit-level perturbations;
and \textbf{RA-BNN}~\cite{he2020defending}, which extends binary neural networks with robust activation functions and training procedures to enhance resilience against parameter-space attacks. We tried to include BITSHIELD~\cite{chen2025bitshield}, but its implementation is not publicly available.
For a fair comparison, we evaluate all methods on the same models and datasets: VGG16 and ResNet32 on CIFAR-10, CIFAR-100, and Tiny-ImageNet. 

\noindent\textbf{Attack Settings.}
We evaluate robustness under three representative parameter attack categories as defined in Section~\ref{sec:threat-model}:  
(1) {Sparse ($L_0$) perturbations}: we retain \textit{ProFlip}~\cite{chen2021proflip}, a strong gradient-based targeted bit-flip attack, as our representative BFA due to its high effectiveness and efficient search;  
(2) {Structured perturbations}: we implement \textit{P3A}~\cite{yao2020deephammer}, which adaptively selects vulnerable channels and perturbs them in a correlated manner;  
(3) {Continuous ($L_2$/$L_\infty$) perturbations}: we adopt \textit{APA}~\cite{rakin2019bit}, an adversarial parameter attack that perturbs a large portion of parameters under norm constraints, using a projected gradient descent~(PGD) optimizer to maximize loss.  
For each attack type, we also design an {adaptive} variant aware of \tool's defense modules.

\noindent\textbf{Attack Details and Budgets.}
We adopt the standard settings from prior parameter-attack works~\cite{chen2021proflip, yao2020deephammer, rakin2019bit, wang2023aegis}.
For ProFlip, we follow the standard ProFlip configuration~\cite{chen2021proflip} with a 200-step search procedure, selecting the target class randomly and estimating candidate bit impact using $k$ sampled candidates ($k{=}20$ for CIFAR-10/100, $k{=}10$ for Tiny-ImageNet), and vary the flip budget $B_f$ in $\{25, 50, 75, 100\}$.

For P3A, we perturb the top gradient-ranked channels with quantization-aware projection, varying the number of perturbed channels $\{1, 3, 5, 7\}$.
For APA, we use a quantization-aware setting in which each PGD update is immediately projected onto the 8-bit quantization grid, matching the deployed parameter representation under \tool. We sweep the $L_\infty$ perturbation magnitude $ \{{\frac{1}{255}, \frac{2}{255}, \frac{4}{255}, \frac{8}{255}, \frac{16}{255}}\}$. This setting evaluates a defense-aware adversary that optimizes over the same quantized parameter manifold as the deployed model.

\begin{table}[t!]
\centering
\setlength{\tabcolsep}{9.9pt}
\footnotesize
\caption{$\tau$ Setting.} 
\label{tab:tau-stats}
\renewcommand{\arraystretch}{1.0}
\begin{tabular}{c|c|c|c|c}
\hline\hline
\textbf{Dataset} & \textbf{Model} & Mean ($\mu$) & Std ($\sigma$) & $\tau$ \\
\hline
\multirow{2}{*}{CIFAR-10}   & ResNet32 & 0.69 & 0.18 & 0.46 \\
           & VGG16    & 0.64 & 0.21 & 0.37 \\
\hline
\multirow{2}{*}{CIFAR-100}  & ResNet32 & 0.43 & 0.16 & 0.23 \\
           & VGG16    & 0.32 & 0.16 & 0.12 \\
\hline
\multirow{2}{*}{Tiny-ImageNet} & ResNet32 & 0.27 & 0.14 & 0.09 \\
              & VGG16    & 0.23 & 0.13 & 0.06 \\
\hline\hline
\end{tabular}
\vspace{-4mm}
\end{table}

\noindent\textbf{Configuration.}
Unless otherwise stated, we use the following default hyperparameter settings for \tool. 
During preparation, we quantize each tensor with uniform 8-bit affine quantization and pack the indices into blocks encoded by a QC-LDPC code with rate $0.875$. We use LDPC blocks of size $N_b=128$, following standard QC-LDPC constructions.
During inference, ARI injects small Gaussian noise into model parameters for stochastic smoothing with $\sigma_w = 10^{-4}$. 
We set the number of stochastic passes to $M_s = 5$ for the fast path and $M_\ell = 25$ for the escalated slow path.
The confidence threshold $\tau$ is determined on clean validation data using a Gaussian approximation of the confidence-margin distribution $\Delta(x)$. 
We estimate the mean $\mu$ and standard deviation $\sigma$ of $\Delta(x)$ and set
\(
\tau = \mu - 1.28\sigma .
\)
This choice corresponds to the nominal 10th-percentile cutoff under a normality assumption and provides a simple, reproducible criterion across datasets and architectures. 
The resulting $\tau$ values are summarized in Table~\ref{tab:tau-stats}.
All experiments are repeated with five independent random seeds affecting defense preparation (KCR keys and LDPC interleaving), attack initialization, and ARI stochastic sampling; we report mean results across runs.
We later examine the impact of \tool\ under variations of these hyperparameters in our sensitivity analysis.

\noindent\textbf{Environment.}
All experiments are performed on a Ubuntu 24.04.3 LTS system with an AMD Ryzen Threadripper PRO 7985WX (64-core) CPU, 502 GiB of DDR5 memory, and NVIDIA RTX 6000 Ada Generation GPU.

\subsection{Metrics}

Following previous works~\cite{wang2023aegis, he2020defending}, we use the following metrics to evaluate model utility, robustness, and deployment overhead.

\noindent \textbf{Clean Accuracy (ACC).} Accuracy on test data without any attack, measuring the utility loss introduced by defenses.

\noindent \textbf{Attack Success Rate (ASR).} The percentage of originally correctly classified samples that are misclassified after an attack. A lower ASR indicates a stronger defense. To ensure comparability across defenses with different clean accuracies, ASR is computed on each model's correctly classified clean subset under the corresponding defense.

\noindent \textbf{Model Size.} On-disk model size (MB) including all auxiliary metadata, together with the percentage increase over {Base}. 

\noindent \textbf{Latency}: Per-query inference latency under identical hardware, batch, and runtime, reported as change to {Base} at P50 and P90, where P50 is the median per-query latency and P90 captures the slower tail (90th percentile) of the distribution. \bin{We also report worst-case latency when all inputs are escalated to ARI's slow path.}

\begin{table}[t!]
\setlength{\tabcolsep}{2.9pt}
\renewcommand{\arraystretch}{1.0}
\footnotesize
\centering
\caption{Model ACC influence evaluation.}
\label{tab:acc}
\begin{tabular}{c|c|c|cccc}
\hline\hline
\multirow{2}{*}{\textbf{Dataset}} & \multirow{2}{*}{\textbf{Model}} & \multirow{2}{*}{\textbf{\begin{tabular}{c}Base\\ACC (\%)\end{tabular}}} & \multicolumn{4}{c}{\textbf{$\Delta$ ACC (\%)}} \\
\cline{4-7} 
 & &    & BIN & RA-BNN & Aegis  & \tool \\
\hline
\multirow{2}{*}{CIFAR-10}   & ResNet32 & 92.88 & -10.17 & -7.67  & -17.20 & \textbf{-1.64} \\
                            & VGG16    & 93.57 & -7.44  & -3.30  & -6.67  & \textbf{-2.08} \\
\hline
\multirow{2}{*}{CIFAR-100}  & ResNet32 & 67.02  & -14.97 & -10.69 & -11.96 & \textbf{-4.90} \\
                            & VGG16    & 72.36  & -24.72 & -7.07  & -11.35 & \textbf{-5.27} \\
\hline
\multirow{2}{*}{\begin{tabular}{c}Tiny\\-ImageNet\end{tabular}} & ResNet32 & 53.93  & -9.29 & -17.94 & -4.73  &  \textbf{-2.48} \\
                               & VGG16    & 61.27 & -12.33 & -7.75  & -8.65  & \textbf{-3.86} \\
\hline\hline
\end{tabular}
\vspace{-4mm}
\end{table}

\section{Results and Analysis}

\begin{figure*}[t!]
\centering
\captionsetup[subfigure]{justification=centering}

\begin{subfigure}[t]{0.333\linewidth}
  \centering
  \includegraphics[width=\linewidth]{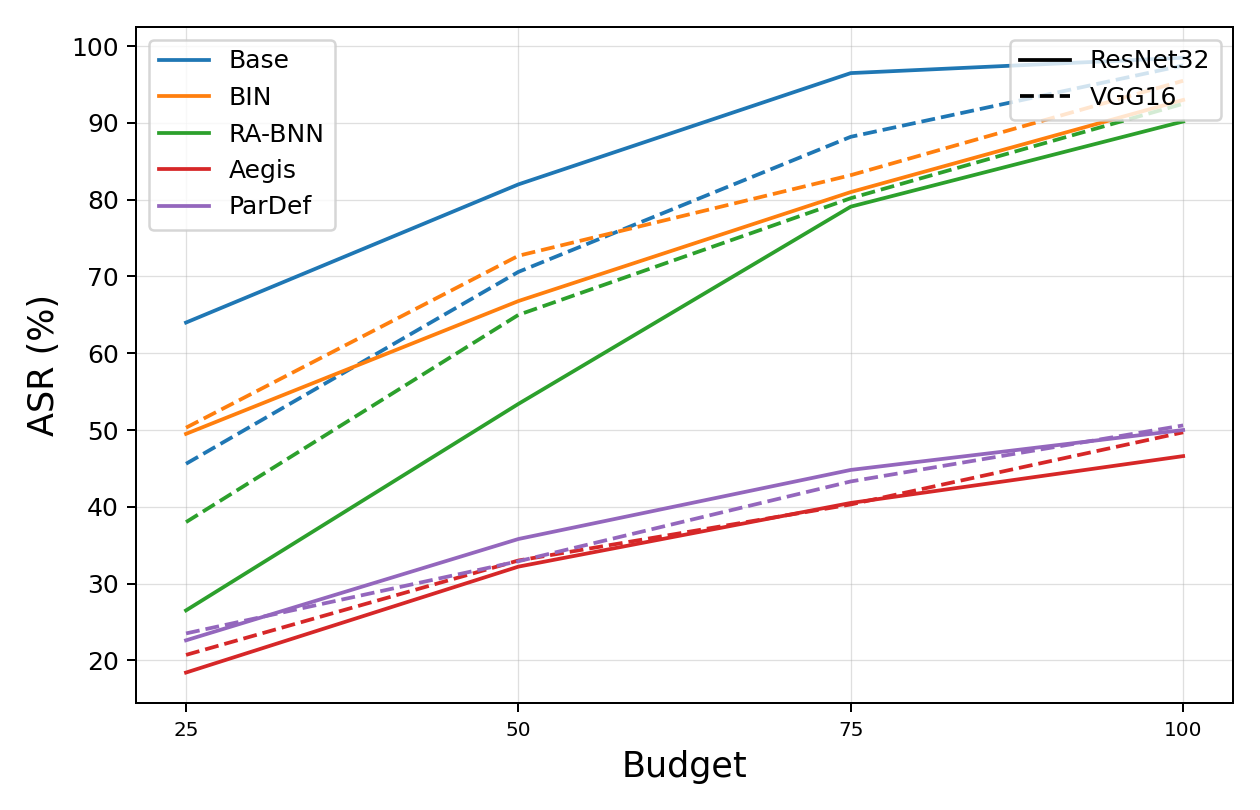}
  \caption{ProFlip — CIFAR-10}
\end{subfigure}\hfill
\begin{subfigure}[t]{0.333\linewidth}
  \centering
  \includegraphics[width=\linewidth]{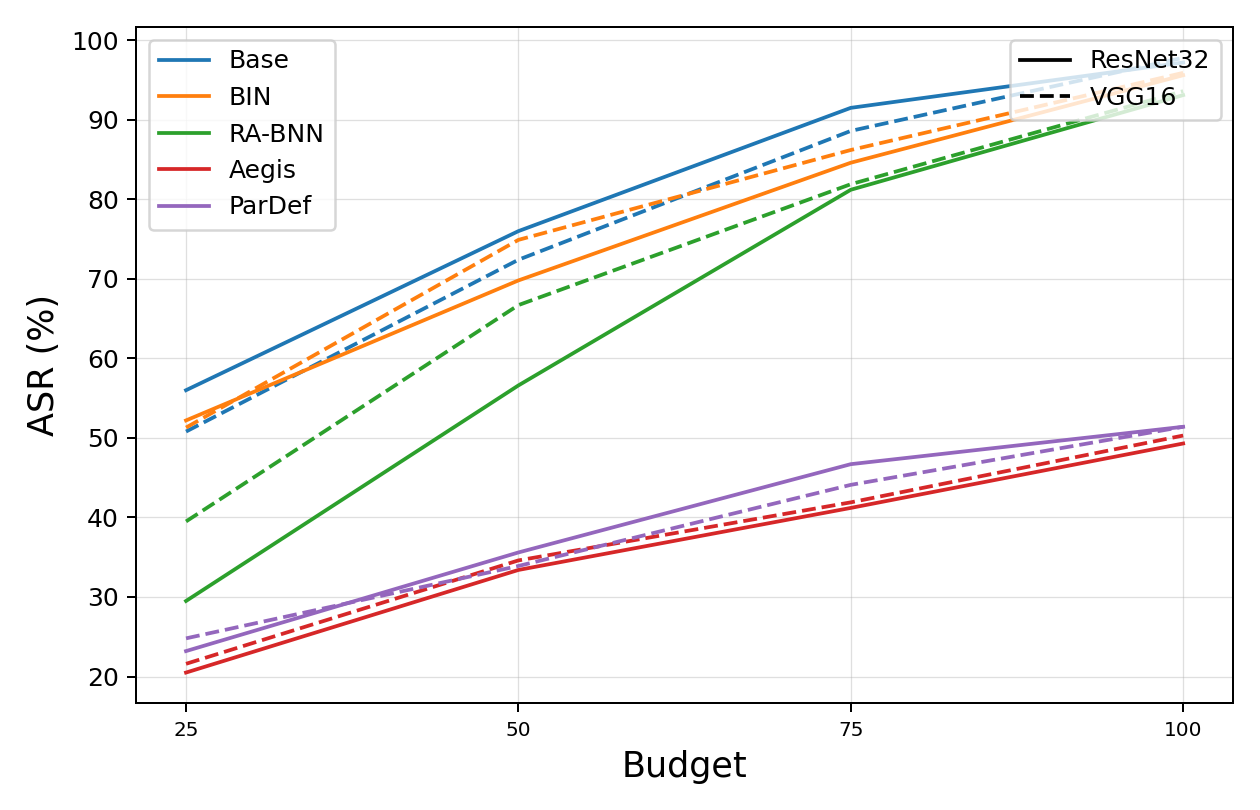}
  \caption{ProFlip — CIFAR-100}
\end{subfigure}\hfill
\begin{subfigure}[t]{0.333\linewidth}
  \centering
  \includegraphics[width=\linewidth]{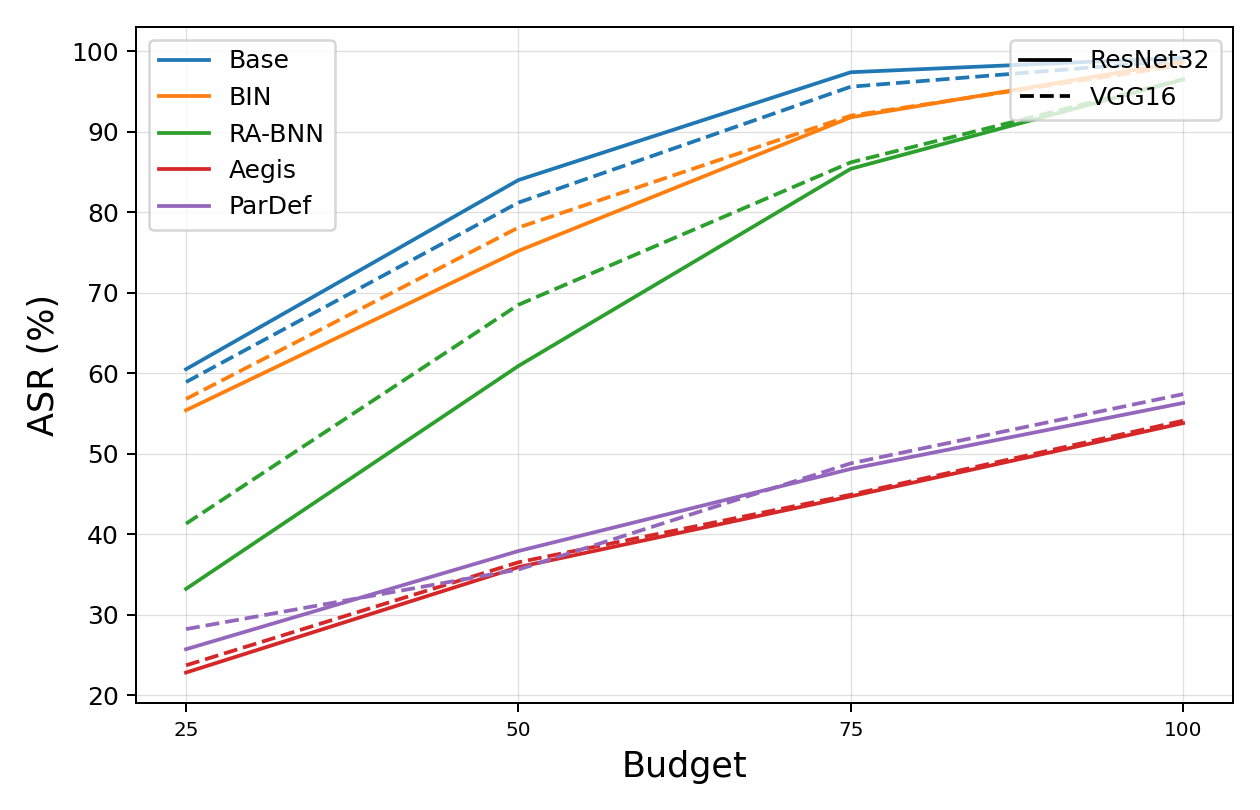}
  \caption{ProFlip — Tiny-ImageNet}
\end{subfigure}

\begin{subfigure}[t]{0.333\linewidth}
  \centering
  \includegraphics[width=\linewidth]{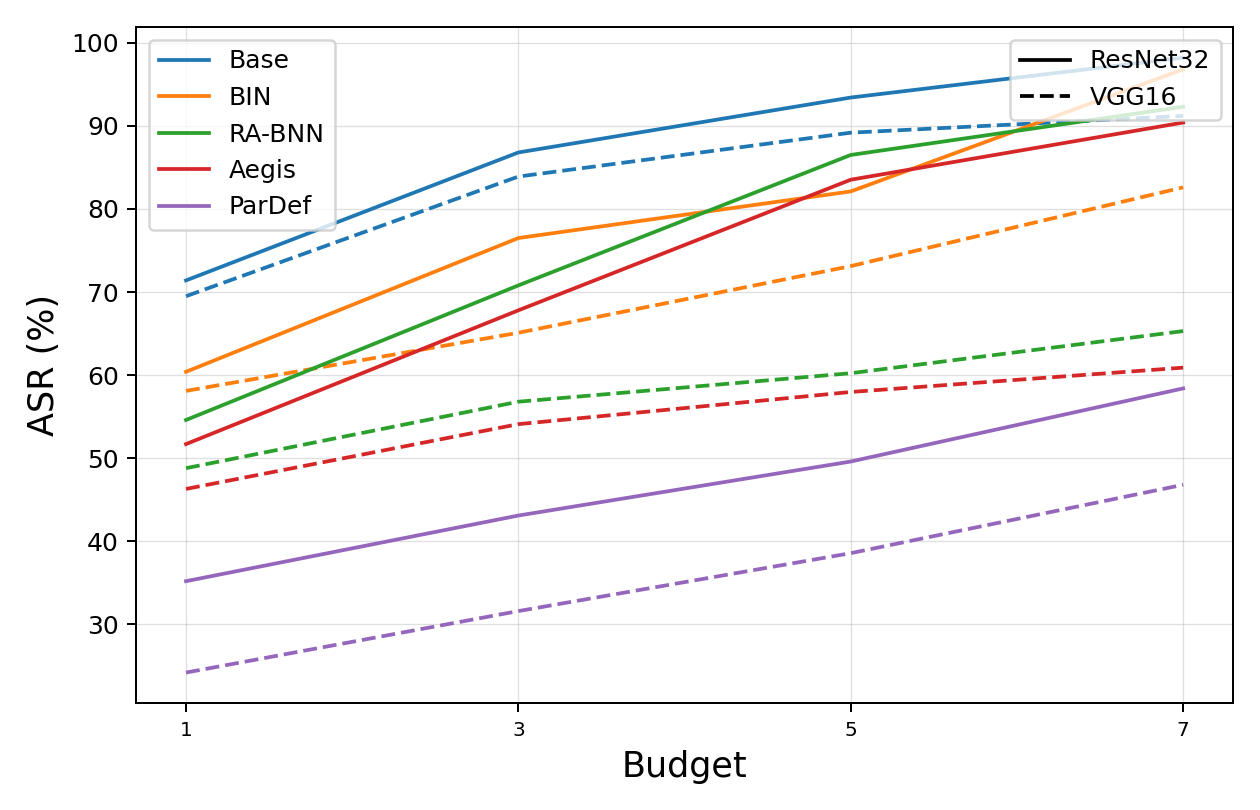}
  \caption{P3A — CIFAR-10}
\end{subfigure}\hfill
\begin{subfigure}[t]{0.333\linewidth}
  \centering
  \includegraphics[width=\linewidth]{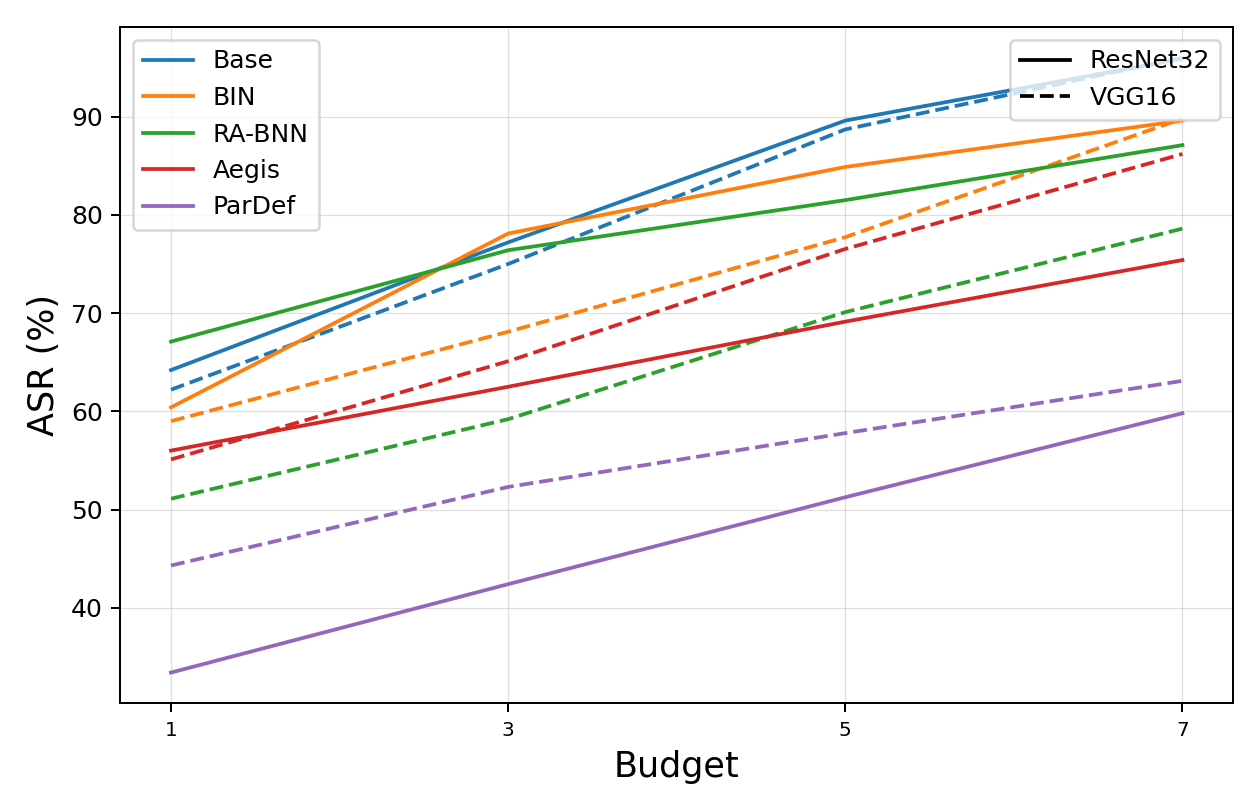}
  \caption{P3A — CIFAR-100}
\end{subfigure}\hfill
\begin{subfigure}[t]{0.333\linewidth}
  \centering
  \includegraphics[width=\linewidth]{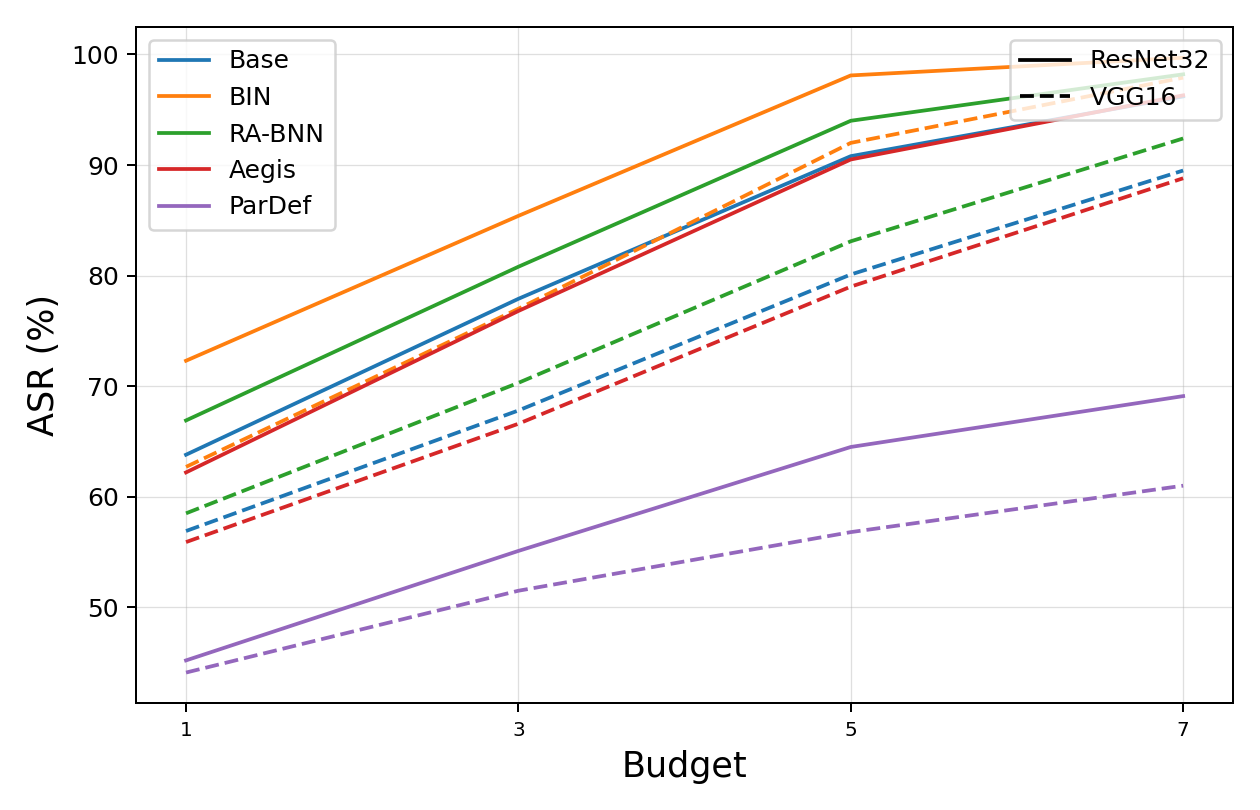}
  \caption{P3A — Tiny-ImageNet}
\end{subfigure}

\begin{subfigure}[t]{0.333\linewidth}
  \centering
  \includegraphics[width=\linewidth]{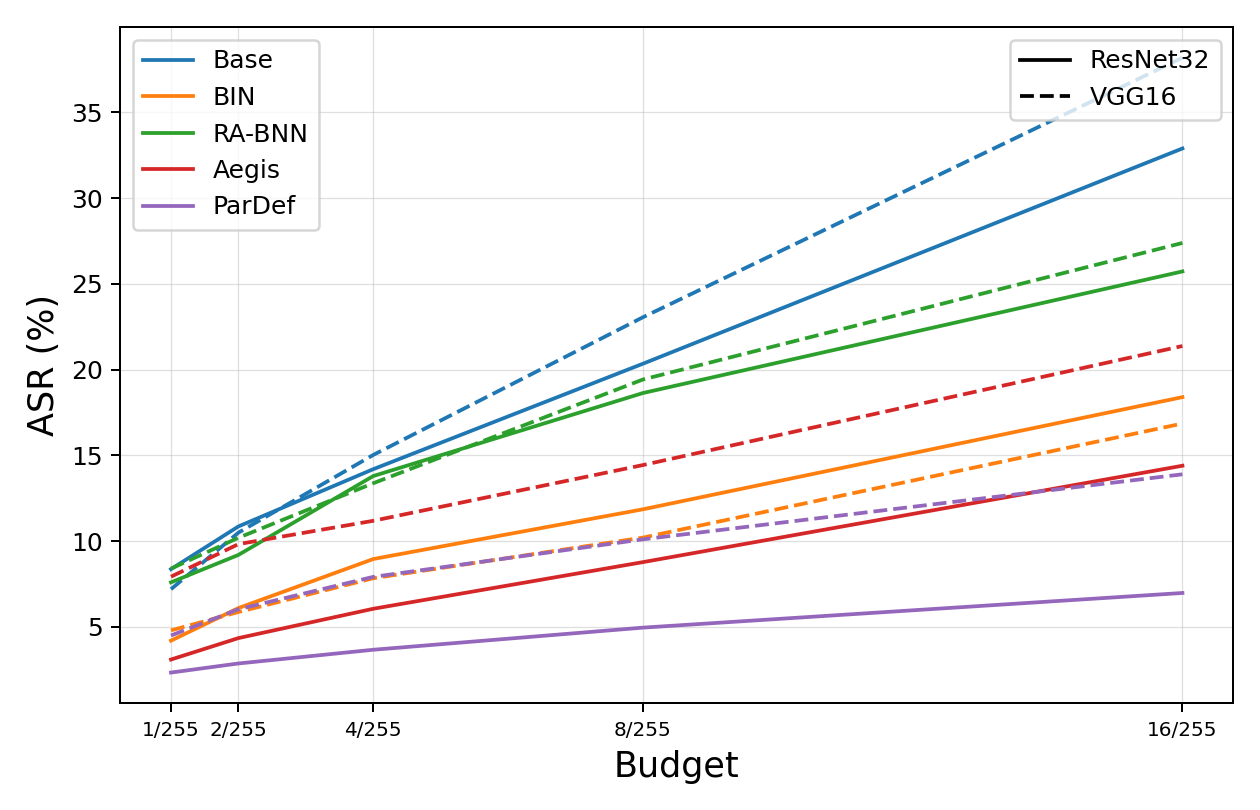}
  \caption{APA — CIFAR-10}
\end{subfigure}\hfill
\begin{subfigure}[t]{0.333\linewidth}
  \centering
  \includegraphics[width=\linewidth]{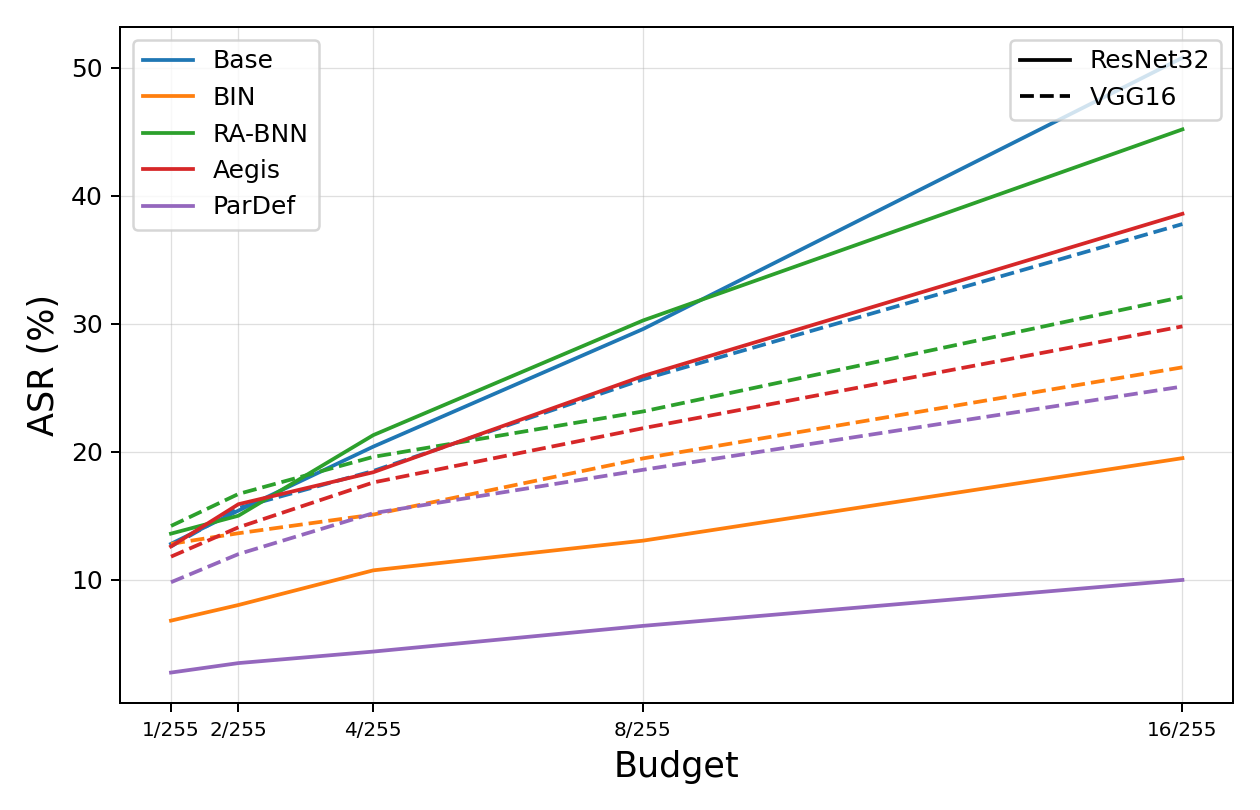}
  \caption{APA — CIFAR-100}
\end{subfigure}\hfill
\begin{subfigure}[t]{0.333\linewidth}
  \centering
  \includegraphics[width=\linewidth]{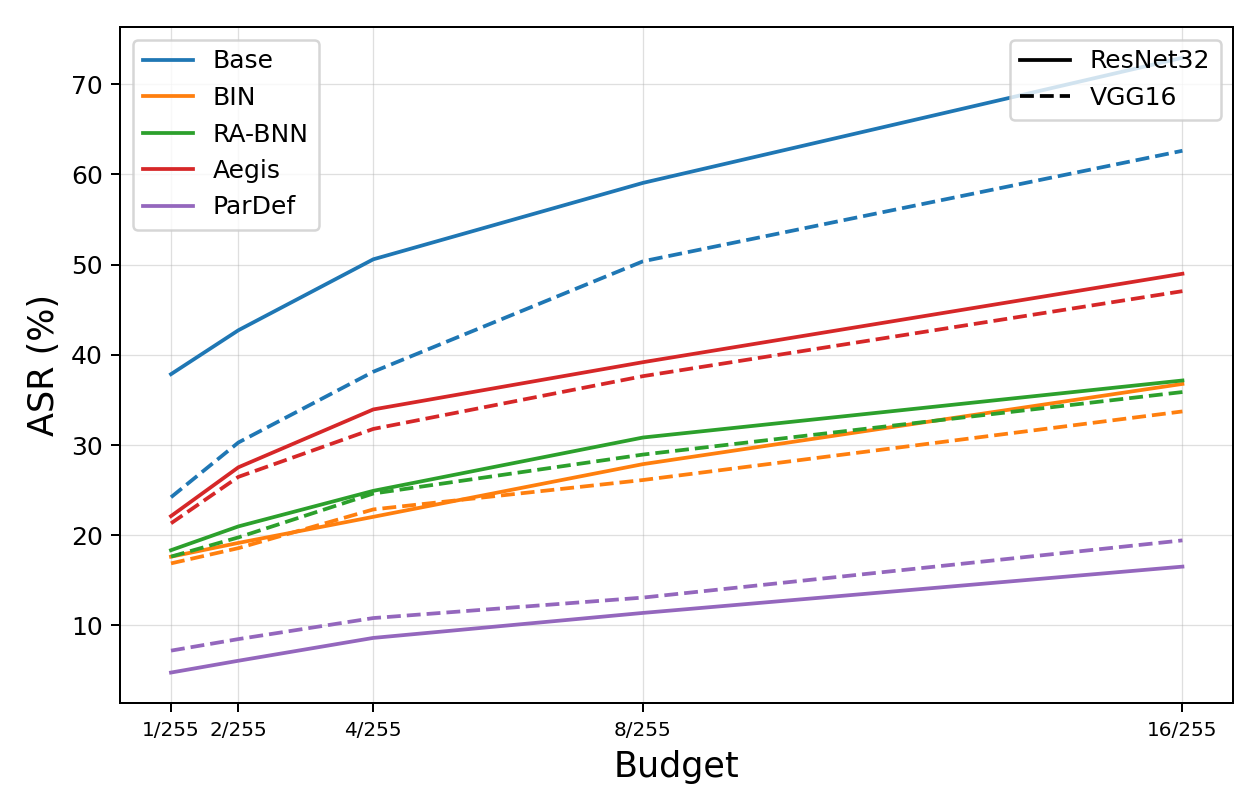}
  \caption{APA — Tiny-ImageNet}
\end{subfigure}
\caption{Comparison of ASR~(\%) across Three Attacks.}
\label{fig:asr_grid_3x3}
\vspace{-4mm}
\end{figure*}

\subsection{RQ1. Model Utility Evaluation}
\label{sec:4.3}

As shown in Table~\ref{tab:acc}, \tool\ achieves the smallest $\Delta$ACC across all datasets and models, indicating strong preservation of clean accuracy. BIN and RA-BNN reduce accuracy more notably due to the constraints of aggressive quantization, often showing drops that are about 2–4$\times$ larger than those of \tool. Aegis, though effective for bit-flip robustness, shows larger drops under our unified evaluation protocol on several benchmarks. \bin{For Aegis, we report reproduced results from its official open-source implementation under our unified evaluation stack. Its reproduced CIFAR-10/ResNet32 accuracy drop is 17.20\%, compared with 1.26\% in the original paper, so we interpret it as a reproduced baseline under our common protocol.}
The stability of \tool\ comes from its parameter-preserving design: it avoids the expressiveness loss of global quantization and the complexity of architectural changes, instead combining reparameterization, coded quantization, and adaptive inference. This selective use of redundancy yields its lower utility cost.

\subsection{RQ2. Mitigating Attacks}
\label{sec:4.4}

\subsubsection{Mitigating Targeted Attacks}

\begin{table}[t!]
\setlength{\tabcolsep}{2.1pt}
\renewcommand{\arraystretch}{1.0}
\footnotesize
\centering
\caption{Adaptive Attack Evaluation on \tool.}
\label{tab:adaptive-audef}
\begin{tabular}{c|c|cc|cc|cc}
\hline\hline
\multirow{2}{*}{\textbf{Dataset}} & \multirow{2}{*}{\textbf{Model}} 
& \multicolumn{2}{c|}{\textbf{ProFlip}} 
& \multicolumn{2}{c|}{\textbf{P3A}} 
& \multicolumn{2}{c}{\textbf{APA}} \\
\cline{3-8}
 &  & \textbf{Base} & \textbf{\tool} & \textbf{Base} & \textbf{\tool} & \textbf{Base} & \textbf{\tool} \\
\hline
\multirow{2}{*}{CIFAR-10}
& ResNet32 & 82.28 & 40.12 & 93.41 & 51.02 & 20.35 & 5.42 \\
& VGG16    & 70.63 & 34.18 & 89.18 & 39.86 & 23.06 & 10.72 \\
\hline
\multirow{2}{*}{CIFAR-100}
& ResNet32 & 76.01 & 39.91 & 89.69 & 52.33 & 29.59 & 6.98 \\
& VGG16    & 72.44 & 35.07 & 88.75 & 59.04 & 25.66 & 19.31 \\
\hline
\multirow{2}{*}{\begin{tabular}{c}Tiny\\-ImageNet\end{tabular}}
& ResNet32 & 84.50 & 41.21 & 90.83 & 61.03 & 59.05 & 12.08 \\
& VGG16    & 81.22 & 36.92 & 80.12 & 57.94 & 50.36 & 13.66 \\
\hline\hline
\end{tabular}
\vspace{-4mm}
\end{table}

\noindent \textbf{\textit{ProFlip.}}
As shown in Fig~\ref{fig:asr_grid_3x3},
\tool\ consistently maintains low ASR across all budgets and datasets. Its curves remain smooth and stable as the attack budget increases, showing reliable protection against sparse parameter corruptions without erratic degradation. Compared with BIN and RA-BNN, which offer only modest relief over the unprotected Base and exhibit visibly larger ASR increases at medium–high budgets, \tool\ achieves stronger robustness. Aegis still attains the lowest ASR overall, reflecting its specialization for targeted bit-flip attacks, but the gap between \tool\ and Aegis remains small and stable across all settings.

\noindent \textbf{\textit{P3A.}}
 Under structured attacks, in Fig~\ref{fig:asr_grid_3x3}, \tool\ achieves the lowest ASR across all datasets and budgets. Its advantage becomes more evident at higher budgets, and the curves remain smooth and consistent across datasets, showing stable protection against coordinated parameter manipulations. Baselines often fail to limit ASR growth; in several cases, their curves even rise above the unprotected Base at medium to high budgets, indicating that aggressive quantization alone cannot handle structured perturbations. \tool\ maintains a steady advantage, reflecting its ability to disrupt structural correlations in the parameter space and stabilize predictions under correlated parameter attacks.

\noindent \textbf{\textit{APA.}}
For APA, \tool\ shows the lowest ASR across all datasets and budgets in Fig~\ref{fig:asr_grid_3x3}, with a clear margin over the other defenses. This margin persists under all budgets, indicating that its stochastic smoothing and confidence-gated voting effectively counter the cumulative effect of continuous parameter drift. BIN and RA-BNN provide some improvement over the unprotected Base, while Aegis performs competitively in certain cases but is often less stable, with higher ASR than quantization-based defenses on several benchmarks.

\noindent \textbf{\textit{Summary.}}
These results highlight the generality of \tool: while Aegis is tailored for bit-flip attacks, \tool\ maintains low and stable ASR under both structured and continuous attacks across datasets and budgets. Its parameter-preserving design does not require retraining, making it applicable and suitable for deployment.

\subsubsection{Mitigating Adaptive Attacks}

To evaluate robustness under a defense-aware adversary and avoid optimistic bias from non-adaptive baselines, we simulate a reconnaissance-then-attack protocol. The reconnaissance phase is kept short not as a security boundary, but as a standardized analysis window to emulate the attacker’s initial profiling stage. \tool\ does not rely on limiting the number of attacker queries: an adversary with unlimited local access to the transformed model can still perform arbitrary offline analysis, yet \bin{an adversary with unlimited local access to the transformed model can still perform arbitrary offline analysis, but without access to the protected KCR keys, exact recovery of the original per-layer channel order remains difficult under the evaluated probing strategies and budgets.}

\noindent \textbf{\textit{Adaptive ProFlip.}}
We estimate bit saliency under ARI via EOT with 20 parameter--noise samples ($\sigma_w{=}10^{-4}$), and then concentrate flips within a small number of QC-LDPC blocks to stress load-time decoding. Concretely, we allocate the same total flip budget~(50) but prioritize flipping higher-impact bits within the same blocks, aiming to trigger \emph{detected-but-uncorrectable} outcomes (e.g., decoder non-convergence or failed parity checks) rather than relying on any fixed bounded-distance correction radius assumption.

\noindent \textbf{\textit{Adaptive P3A.}}
Run perturbation–observation cycles on the top-10 most sensitive channels per layer, using BN parameter analysis to infer permutation-consistent channel groups. Apply structured perturbations to the inferred channel clusters, targeting 5 channels distributed across vulnerable layers under the original P3A budget. Project all updates to the 8-bit quantization grid after each perturbation step.

\noindent \textbf{\textit{Adaptive APA.}}
Use EOT–PGD in parameter space: at each step, sample 5 noisy parameter realizations and average gradients to minimize the confidence margin $\Delta(x)$, aiming to trigger ARI's escalation mechanism and bias the voting outcome. Project updates to both the 8-bit quantization grid and the original constraint ($8/255$).

\noindent \textbf{\textit{Results.}}
As shown in Table~\ref{tab:adaptive-audef}, even with defense-aware tactics at equal budgets, \tool\ prevents collapse and preserves strong, stable robustness.
For ProFlip, block-concentrated flips weaken LDPC, yet most errors are detected rather than silently miscorrected, and KCR diffuses bit effects, so the ASR lift is bounded.
For P3A, channel probing partially recovers structure and trims KCR’s advantage, but permutation/BN constraints and 8-bit projection limit alignment, and \tool\ retains a clear margin.
For APA, EOT-PGD improves consistency under ARI, but smoothing and confidence-gated voting suppress flips.
Across all adaptive attacks, \tool\ reduces ASR by about 40–70\% compared with Base, showing that its robustness does not collapse even under defense-aware adversaries.

\bin{
\subsubsection{Mitigating Joint Adaptive Attacks}
\label{sec:joint-attack}
To evaluate joint defense-aware attacks, we construct three joint adaptive attacks.

\noindent \textbf{\textit{Joint-ProFlip.}}
Joint-ProFlip builds on Adaptive ProFlip by additionally introducing KCR-aware channel targeting. 
The attacker uses channel-wise deployment statistics to approximate sensitive channel groups, then applies the same flip budget ($B_f=50$), QC-LDPC block concentration, and ARI-aware EOT saliency used in Adaptive ProFlip.

\noindent \textbf{\textit{Joint-P3A.}}
Joint-P3A strengthens Adaptive P3A by combining KCR-aware channel-group probing with ARI-aware EOT gradients. 
Following Adaptive P3A, the attacker ranks the top-10 sensitive channels per layer, perturbs 5 channels across vulnerable layers under the original P3A budget, projects updates to the deployed 8-bit quantization grid, and averages gradients over 5 ARI noise samples.

\noindent \textbf{\textit{Joint-APA.}}
Joint-APA strengthens Adaptive APA by adding KCR-aware sensitivity weighting on the transformed deployed model. 
Following Adaptive APA, each PGD step averages gradients over 5 ARI noise samples, optimizes the attack objective with a confidence-margin reduction term, and projects the update to both the $L_\infty$ budget ($8/255$) and the deployed 8-bit quantization grid.

\noindent\textbf{\textit{Results.}}
Table~\ref{tab:joint-attack} shows that the joint adaptive attacks consistently increase ASR compared with the corresponding adaptive variants in Table~\ref{tab:adaptive-audef}. This is expected because the joint attacks no longer adapt to a single defense behavior in isolation; instead, they combine KCR-aware localization, quantization-aware perturbation, and ARI-aware EOT optimization, with Joint-ProFlip further incorporating QC-LDPC block-aware concentration. Joint-P3A causes the largest increase, with ASR rising by 7.97--10.81 percentage points, suggesting that structured attacks benefit most from partially recovering channel-group information under KCR. Joint-ProFlip increases ASR by 6.46--8.83 percentage points, indicating that concentrating bit flips within vulnerable coded blocks makes sparse corruption more effective, but QC-LDPC and KCR still prevent a collapse. Joint-APA shows a smaller increase of 4.34--6.61 percentage points, implying that ARI's stochastic smoothing and quantization-aware projection continue to limit dense parameter drift. Overall, although the joint attacks are stronger than the module-specific adaptive attacks, their ASR remains below the corresponding Base results in all settings and is substantially lower in most cases. This indicates that \tool's robustness does not rely on attackers adapting to only one module at a time.

\begin{table}[t]
\setlength{\tabcolsep}{4pt}
\renewcommand{\arraystretch}{1.0}
\footnotesize
\centering
\caption{Joint Adaptive Attack Evaluation on \tool.}
\label{tab:joint-attack}
\begin{tabular}{c|c|c|c|c}
\hline\hline
\multirow{2}{*}{\textbf{Dataset}} & \multirow{2}{*}{\textbf{Model}} 
& \multicolumn{3}{c}{\textbf{ASR (\%)}} \\
\cline{3-5}
& & \textbf{Joint-ProFlip} & \textbf{Joint-P3A} & \textbf{Joint-APA} \\
\hline
\multirow{2}{*}{CIFAR-10}
& ResNet32 & 48.13 & 60.47 & 9.76 \\
& VGG16    & 41.92 & 49.31 & 15.08 \\
\hline
\multirow{2}{*}{CIFAR-100}
& ResNet32 & 46.37 & 63.14 & 11.69 \\
& VGG16    & 43.21 & 67.82 & 24.36 \\
\hline
\multirow{2}{*}{Tiny-ImageNet}
& ResNet32 & 50.04 & 69.73 & 18.21 \\
& VGG16    & 44.68 & 65.91 & 20.27 \\
\hline\hline
\end{tabular}
\vspace{-3mm}
\end{table}

}

\subsection{RQ3. Deployment Overhead}
\label{sec:4.5}

Table~\ref{tab:overhead} indicates two deployment aspects, model size~(MB) and inference latency~(ms). 
Across all the selected models and datasets, {\tool\ consistently achieves an approximately 70\% reduction in size compared to Base, primarily because weights are stored in 8-bit quantized form; QC-LDPC adds only modest redundancy on top of these indices. This reduction makes the protected models easier to deploy. For latency, \tool\ introduces modest overhead relative to Base under typical workloads: P50 typically increases by about 0--7\% and P90 by about 4--9\% across datasets, with no pronounced tail amplification as models scale. In short, the results suggest that \tool\ offers a significant reduction in model size with only a small impact on typical-case runtime.}
This latency increase is attributable to ARI's stochastic inference: the fast path determines the small P50 rise, while the slow path accounts for the bounded P90 tail. KCR and QC-LDPC are executed once at model loading and incur no per-query overhead. Thus, the observed runtime cost reflects ARI's adaptive redundancy rather than preparation-time transformations.
\bin{We additionally report a {worst-case} latency column, corresponding to the scenario where every query is escalated to the slow path ($M_\ell = 25$ passes), e.g., under a denial-of-service-style adversary. Because ARI's stochastic passes are executed as a single batched forward pass on the GPU, the actual slowdown relative to P50 is approximately $3.5\times$--$4\times$ rather than the naive $M_\ell/M_s = 5\times$. Deployment-level mitigations (e.g., rate-limiting, capping $M_\ell$ under sustained escalation) can further bound this overhead if needed.}

\begin{table}[t!]
\setlength{\tabcolsep}{2pt}
\renewcommand{\arraystretch}{1.0}
\footnotesize
\centering
\caption{Deployment Overhead.}
\label{tab:overhead}
\begin{tabular}{c|c|cc|cc|c}
\hline\hline
\multirow{2}{*}{\textbf{Dataset}} & \multirow{2}{*}{\textbf{Model}} 
& \multicolumn{2}{c|}{\textbf{Model Size}} 
& \multicolumn{2}{c|}{\textbf{Latency P50/P90}} 
& \bin{\textbf{Worst}} \\
\cline{3-7}
 &  & Base & \tool & Base & \tool & \bin{\tool} \\
\hline
\multirow{2}{*}{CIFAR-10}   
 & ResNet32 & 1.94 & \textbf{0.57} & \textbf{1.29/1.67} & 1.29/1.74 & \bin{4.84} \\
 & VGG16    & 58.92 & \textbf{17.33} & \textbf{2.85/3.76} & 3.00/4.03 & \bin{11.04} \\
\hline
\multirow{2}{*}{CIFAR-100}  
 & ResNet32 & 1.95 & \textbf{0.57} & \textbf{1.25/1.74} & 1.34/1.85 & \bin{5.25} \\
 & VGG16    & 60.06 & \textbf{17.66} & \textbf{2.91/3.85} & 3.10/4.14 & \bin{11.81} \\
\hline
\multirow{2}{*}{\begin{tabular}{c}Tiny\\-ImageNet\end{tabular}} 
 & ResNet32 & 1.98 & \textbf{0.58} & \textbf{4.86/6.48} & 5.14/6.98 & \bin{19.84} \\
 & VGG16    & 60.48 & \textbf{17.78} & \textbf{10.80/14.21} & 11.56/15.48 & \bin{41.85} \\
\hline\hline
\end{tabular}
\vspace{-3mm}
\end{table}

\begin{table}[t!]
\setlength{\tabcolsep}{0.5pt}
\renewcommand{\arraystretch}{1.0}
\footnotesize
\centering
\caption{Ablation on Different Module of \tool.}
\label{tab:ablation-main}
\begin{tabular}{l|c|c|c|c|c|c}
\hline\hline
\textbf{Variant} &
\textbf{ProFlip} &
\textbf{P3A} &
\textbf{APA } &
\textbf{$\Delta$ACC(\%)} &
\textbf{Model Size} &
\textbf{P50/P90} \\
\hline
\tool\ & \textbf{22.1} & \textbf{18.4} & \textbf{15.7} & -1.64 & \textbf{0.57} & 1.29 / 1.74 \\
\hline
w/o KCR              & 39.8          & 67.6          & 33.4          & -1.62 & 0.57 & 1.27 / 1.72 \\
w/o QC-LDPC          & 68.9          & 34.1          & 29.6          & -0.96 & 2.11 & 1.46 / 2.00 \\
w/o ARI              & 41.2          & 36.7          & 42.3          & {-0.72} & 0.57 & {1.20 / 1.60} \\
\hline
\bin{w/o KCR+QC-LDPC}      & 77.8          & 74.6          & 37.2          & -0.38 & 1.94 & 1.45 / 1.98 \\
\bin{w/o KCR+ARI   }       & 56.4          & 82.7          & 50.8          & -0.68 & 0.57 & \textbf{1.18 / 1.56} \\
\bin{w/o QC-LDPC+ARI}      & 74.9          & 46.7          & 53.4          & \textbf{-0.03} & 2.11 & 1.29 / 1.67 \\
\hline\hline
\end{tabular}
\vspace{-4mm}
\end{table}

\subsection{RQ4. Ablation Study}
\label{sec:4.6}
\bin{Here we ablate \tool\ on CIFAR-10/ResNet32 using both single-module and combined-module variants.
The single-module ablation quantifies the individual contribution of KCR, QC-LDPC coded quantization, and ARI, while the combined ablation examines whether the robustness gains arise from complementary interactions among modules rather than from one dominant component alone.}
As shown in Table~\ref{tab:ablation-main}, removing KCR causes the largest degradation under structured perturbations and also increases ASR under sparse and continuous attacks. This aligns with KCR's role: keyed channel permutations disrupt consistent channel or group alignments and diffuse localized errors, so taking it away restores exploitable structure.
Removing QC-LDPC primarily harms robustness to sparse attacks while only moderately affecting P3A and APA. In our pipeline, the coded-quantization stage also underpins compact storage, so dropping it forfeits that compact representation and increases both model size and per-query latency, consistent with operating on a higher-precision checkpoint and the resulting memory/compute overheads during inference.
Removing ARI most strongly degrades resilience to dense, small-magnitude parameter noise and also weakens tolerance to structured perturbations. Because ARI adds stochastic passes only for low-margin inputs, disabling it slightly improves clean accuracy and reduces median or tail latency, illustrating the intended robustness–utility trade-off of the adaptive inference path.
\bin{
The combined-module ablation reveals clear module-specific trade-offs. 
The w/o KCR+QC-LDPC variant, which retains only ARI, better mitigates APA than ProFlip/P3A, but loses QC-LDPC's compression benefit, leading to a larger model size (1.94MB) and higher latency (1.45/1.98). 
The w/o KCR+ARI variant, which retains only QC-LDPC, keeps the smallest model size (0.57MB) and lowest latency (1.18/1.56), and provides more protection against ProFlip than against structured or dense perturbations. 
The w/o QC-LDPC+ARI variant, which retains only KCR, nearly preserves clean accuracy ($\Delta$ACC=-0.03) and helps more on P3A, but without QC-LDPC and ARI it provides weak protection against ProFlip and APA and loses the storage benefit. 
Thus, the full \tool\ provides the best robustness--efficiency balance: QC-LDPC supplies compact storage and bit-level protection, KCR disrupts structured attacks, and ARI stabilizes residual dense perturbations.
}

\section{Discussion}

\subsection{Parameter-Space Sensitivity Analysis}

\begin{figure}[t!]
    \centering
    \begin{subfigure}[b]{\linewidth} %
        \centering
        \includegraphics[width=0.95\textwidth]{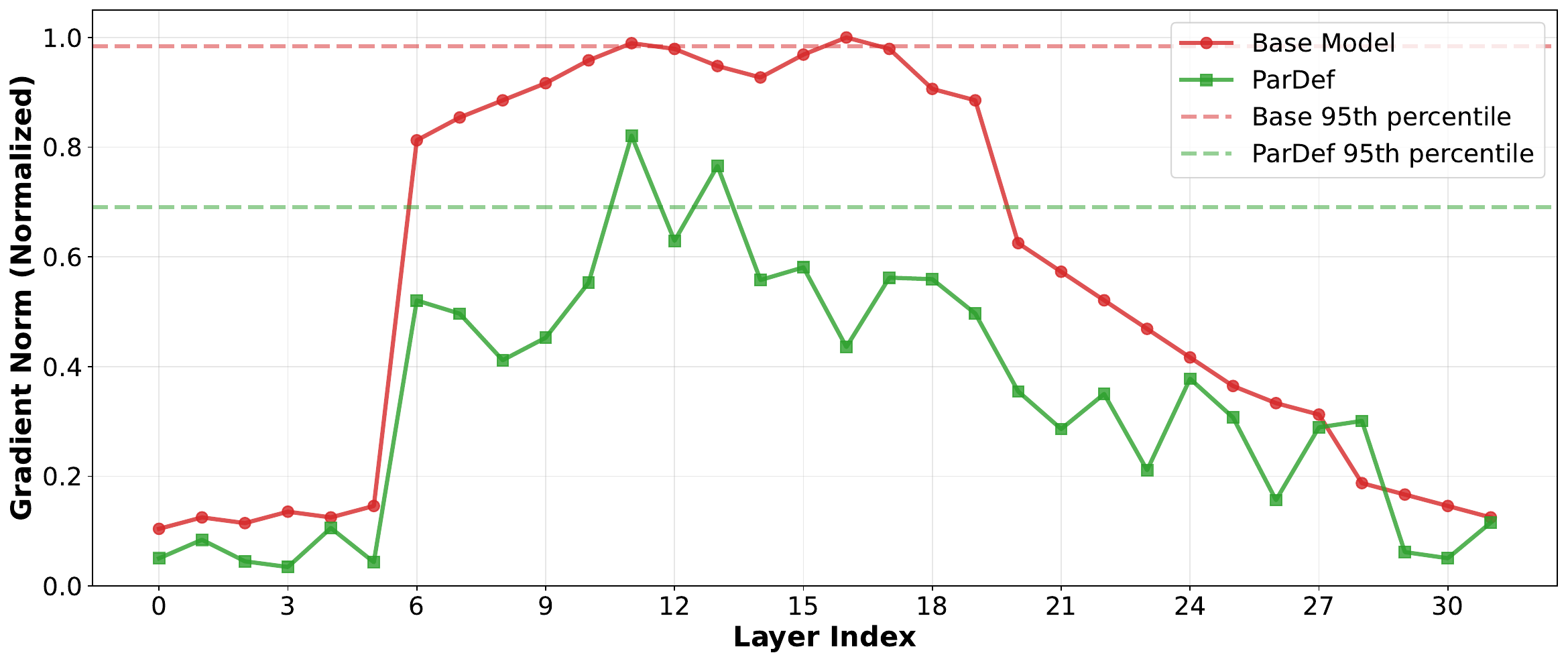}
        \caption{Trend of gradient magnitude.}
        \label{fig:sub:lineplot}
    \end{subfigure}

    \vspace{1mm}
    
    \begin{subfigure}[b]{\linewidth}
        \centering
        \includegraphics[width=0.95\textwidth]{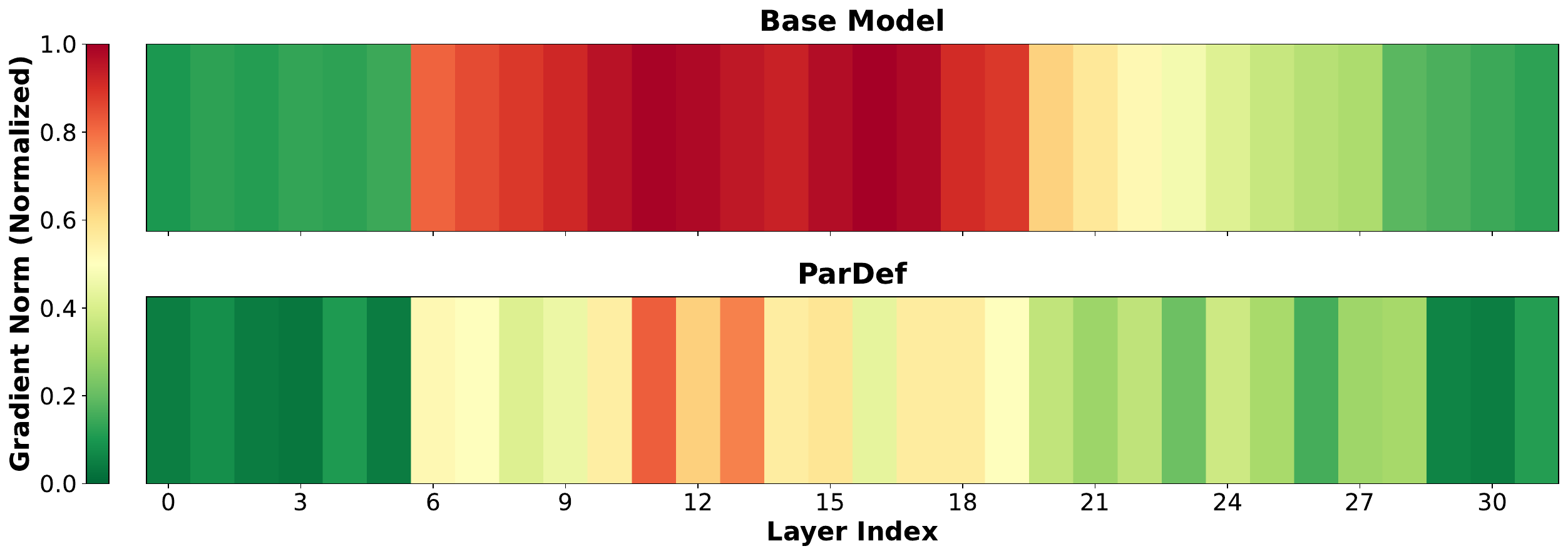}
        \caption{Heatmaps of gradient magnitude.}
        \label{fig:sub:heatmap}
    \end{subfigure}
    \caption{Layer-wise Sensitivity Analysis.} 
    \label{fig:param_sensitivity}
    \vspace{-4mm}
\end{figure}

\begin{figure*}[t!]
\centering
\includegraphics[width=1\textwidth]{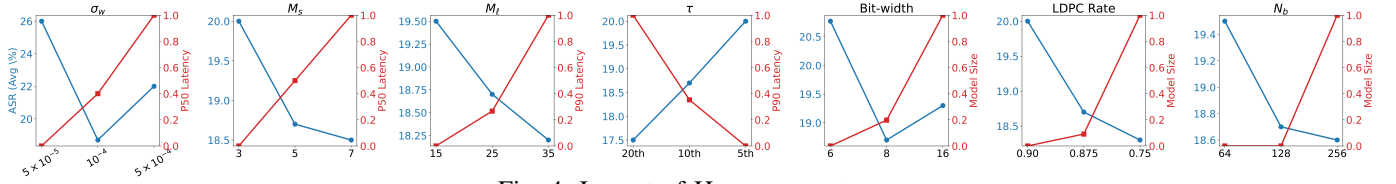}
\vspace{-7mm}
\caption{Impact of Hyperparameters.}
\vspace{-4mm}
\label{fig:hyper}
\end{figure*}

Figure~\ref{fig:param_sensitivity} reports the layer-wise gradient norms of ResNet32/CIFAR-10 with respect to the loss under small parameter perturbations. The base model exhibits pronounced vulnerability concentration: several middle layers (10–18) show sharp gradient spikes near the 95th percentile, indicating that small weight parameter deviations in these layers induce large output changes and thus form natural attack hotspots. After applying \tool, both the line trend and the heatmaps reveal a substantial flattening of these peaks: the high-sensitivity layers shrink in magnitude, and the overall distribution becomes smoother and more uniform across the network. This reduction of localized sharp gradients reflects the combined effect of KCR (diffusing layer-wise influence), QC-LDPC (forcing weights onto stable quantization manifolds), and ARI (attenuating prediction variance), together yielding a parameter landscape where parameter-oriented attacks produce far less amplification, explaining the consistent ASR reductions observed across sparse, structured, and continuous attacks.

\subsection{Impact of Hyperparameters in \tool}

We evaluate the sensitivity of seven hyperparameters in \tool\ on ResNet-32/CIFAR-10.  
For stochastic smoothing, we vary the noise scale $\sigma_w\!\in\!\{5\times10^{-5},10^{-4},5\times10^{-4}\}$ and the number of fast-path passes $M_s\!\in\!\{3,5,7\}$; these two mainly affect ASR and P50 latency, while changes in clean ACC and model size can be omitted.
For escalation, we vary the slow-path passes $M_\ell\!\in\!\{15,25,35\}$ and set the confidence threshold $\tau$ to the 5th, 10th, and 20th percentiles of the clean-margin distribution. Together they control how often the slow path is invoked and thus primarily impact ASR and P90 latency.
For QC-LDPC, we test block sizes $N_b\!\in\!\{64,128,256\}$ and coding rates $\{0.75,0.875,0.90\}$, and for post-preparation quantization we use bit-widths $\{6,8,16\}$. These coding and quantization hyperparameters jointly determine the redundancy of stored weights and mainly influence ASR and model size.
Fig~\ref{fig:hyper} evaluates the trade-off between robustness (ASR, lower is better) and deployment overhead (Latency/Model Size) across seven hyperparameters. The analysis confirms that the default settings represent an optimal deployment balance. For the Adaptive Robust Inference (ARI) module, increasing fast-path passes ($M_s$) or the noise scale ($\sigma_w$) beyond defaults yields diminishing returns, providing marginal ASR gains while substantially raising P50 latency. Similarly, the escalation mechanism shows a direct trade-off: tuning the slow-path passes ($M_\ell$) or loosening the confidence threshold ($\tau$) offers slight ASR improvements at the cost of significantly increased P90 latency. For the storage modules, the 8-bit quantization provides the best ASR/Size efficiency. While lower LDPC rates achieve the absolute lowest ASR by increasing redundancy, the default rate of 0.875 maintains strong robustness with a minimized size penalty. In summary, further increasing hyperparameter values yields only marginal ASR gains, often making larger settings computationally and storage-wise inefficient, validating our selection of the balanced default configuration.

\begin{figure*}[t!]
\centering
\includegraphics[width=1\linewidth]{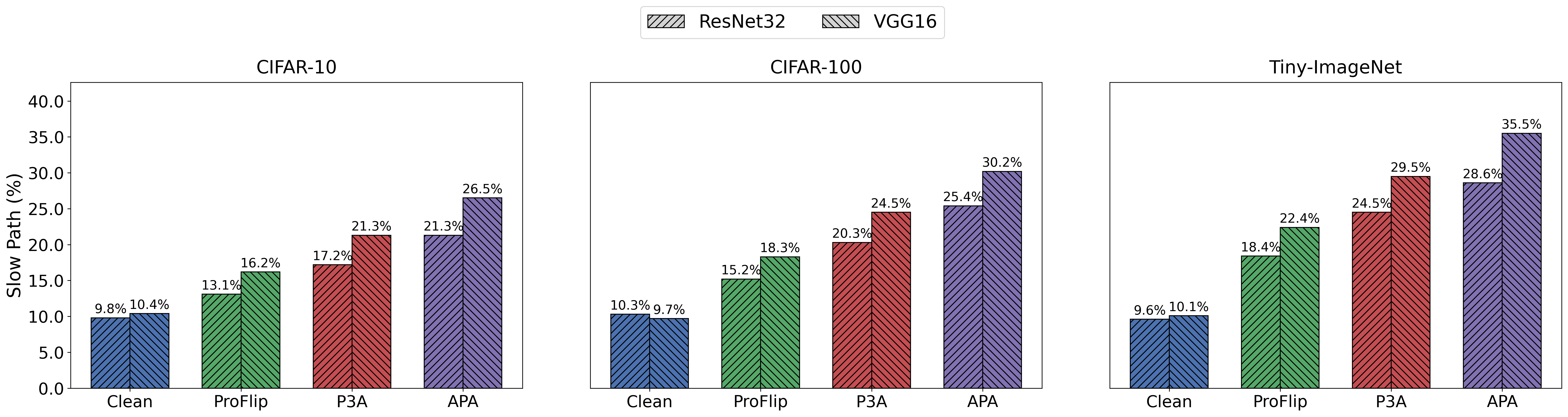}
\vspace{-4mm}
\caption{Slow Path Rate.}
\vspace{-4mm}
\label{fig:slr}
\end{figure*}

\bin{
\subsection{Security Analysis of KCR}

A natural question is whether a white-box adversary can approximate or recover KCR's secret transforms $\{(P_l,Q_l)\}$ through probing or statistical analysis. We do not claim cryptographic secrecy for the transformed parameters; instead, KCR increases the difficulty of mapping deployed channels back to the original attack-sensitive directions when key material is protected. For a layer with $C$ output channels, the permutation space contains $C!$ possibilities (over $10^{89}$ for $C=64$), and cross-layer recovery is further constrained by the edge-consistency conditions in Section~\ref{sec:KCR}. Diagonal scalings also weaken simple magnitude-based matching. Practical heuristics based on channel-wise deployment statistics, such as BatchNorm statistics when available, may recover coarse channel groups but not a canonical ordering within each group. Thus, KCR should be viewed as a key-protected reparameterization mechanism that raises the cost of targeted parameter localization, rather than as a standalone cryptographic defense. Our adaptive evaluation includes EOT and probing-based heuristics, so the observed robustness does not rely on gradient obfuscation.
}

\subsection{LDPC Error-Correction Analysis}

To verify that LDPC coded quantization provides effective protection, we conduct an error-correction experiment. We use proflip~(50 bits) to attack the LDPC-encoded model, and classify each affected block as:
(1) {Corrected, means flipped bits fully repaired}, (2) {Detected, means errors detected but uncorrectable}, and (3) {Silent, means errors decoded to a wrong valid codeword}. 
We report block-level statistics in Table~\ref{tab:ldpc-ecc-table}. 
By examining block-wise results, we find that the defense corrects 65–70\% of corrupted blocks. This is because most corrupted blocks remain decodable by our load-time QC-LDPC decoder under the evaluated flip budgets. Among the remaining blocks, 30–35\% are {detected but uncorrectable}, and Silent remain under 0.6\%. 

\begin{table}[t]
\setlength{\tabcolsep}{8.5pt}
\renewcommand{\arraystretch}{1.0}
\footnotesize
\centering
\caption{Error-Correction Results.}
\label{tab:ldpc-ecc-table}
\begin{tabular}{c|c|c|c|c}
\hline\hline
\textbf{Dataset} & \textbf{Model} 
& \textbf{Corrected} 
& \textbf{Detected} 
& \textbf{Silent} \\
\hline
\multirow{2}{*}{CIFAR-10}
& ResNet32 & 68.1\% & 31.5\% & 0.4\% \\
& VGG16    & 70.4\% & 29.1\% & 0.5\% \\
\hline
\multirow{2}{*}{CIFAR-100}
& ResNet32 & 66.7\% & 32.9\% & 0.4\% \\
& VGG16    & 69.8\% & 29.8\% & 0.4\% \\
\hline
\multirow{2}{*}{Tiny-ImageNet}
& ResNet32 & 63.9\% & 35.6\% & 0.5\% \\
& VGG16    & 65.1\% & 34.3\% & 0.6\% \\
\hline\hline
\end{tabular}
\vspace{-4mm}
\end{table}

\subsection{Activation Behavior of ARI}
Figure~\ref{fig:slr} validates the efficiency of ARI's adaptive execution by analyzing the proportion of inputs escalated to the computational slow path. Our design relies on the trigger rate remaining minimal for benign inputs while increasing significantly under parameter attacks. As shown, the trigger rate on clean inputs remains consistently near the 10\% target. Under all three attack types, the slow path activation escalates. The trigger rate reaches up to 35.5\% for VGG16/Tiny-ImageNet under APA. This substantial increase validates that the confidence margin successfully acts as a parameter-drift sensor, enabling ARI to allocate computational redundancy only when the model's integrity is compromised, thereby justifying the adaptive overhead observed in our deployment analysis.

\bin{
\subsection{Generalization to Transformer Architectures}

\subsubsection{Adapting KCR to Vision Transformers}

The KCR formulation in Section~\ref{sec:KCR} can be extended to linear-layer-dominated architectures under the structural constraints described below. We describe its extension to the key components of Vision Transformer architecture.

\noindent\textbf{Multi-head self-attention.}
A transformer block's attention computes $\mathrm{Attn}(x) = \mathrm{softmax}(QK^\top / \sqrt{d_h}) V$ with $Q = x W_Q$, $K = x W_K$, $V = x W_V$, concatenated across $H$ heads of dimension $d_h$ and projected via $W_O$. To preserve equivalence in attention, we use the following sufficient constraints beyond the generic linear-layer case:
(i)~\emph{Shared head-wise permutation across Q/K/V.} Because $QK^\top$ depends on the inner product between $Q$ and $K$, applying distinct output permutations to $W_Q$ and $W_K$ would yield $QK^\top \rightarrow Q\Pi_Q \Pi_K^\top K^\top$ and break equivalence. As a sufficient design, we therefore tie the output permutations: $\Pi^{(\mathrm{out})}_{W_Q} = \Pi^{(\mathrm{out})}_{W_K} = \Pi^{(\mathrm{out})}_{W_V}$, constrained to permute only within each head's $d_h$-dimensional slice; the resulting permutation is block-diagonal across heads, preserving head-wise independence.
(ii)~\emph{Identity scaling on Q and K.} The softmax is not invariant under per-channel scaling, so we set $D^{(\mathrm{out})}_{W_Q} = D^{(\mathrm{out})}_{W_K} = I$ to preserve the attention logits exactly. Scaling on $W_V$ is permitted as it induces a linear transform on the attention output that can be absorbed by $W_O$'s input scaling.
(iii)~\emph{Edge consistency with $W_O$.} The output projection $W_O$ uses the same head-wise permutation on its input side ($\Pi^{(\mathrm{in})}_{W_O}$ tied to the head-wise permutation above), while its output permutation is free and propagates consistently to the residual stream.

\noindent\textbf{Feed-forward network.}
The MLP block applies $\mathrm{MLP}(x)=\mathrm{GELU}(xW_1+b_1)W_2+b_2$. KCR assigns $(\Pi_1^{(\mathrm{out})}, D_1^{(\mathrm{out})})$ to $W_1$'s output and propagates it to $W_2$'s input via edge consistency ($\Pi_2^{(\mathrm{in})} = \Pi_1^{(\mathrm{out})}$, $D_2^{(\mathrm{in})} = D_1^{(\mathrm{out})}$). Because GELU is element-wise, it commutes with permutations; we set the diagonal scaling to identity across the GELU boundary, consistent with the activation rule in Section~\ref{sec:KCR}.

\noindent\textbf{LayerNorm.}
For permutation-only transforms on the hidden dimension, LayerNorm's normalization statistics $(\mu,\sigma)$ are permutation-invariant. Its per-channel affine parameters absorb the transform as in Equation~\ref{eq:kcr-affines-compact}: $\widetilde{A}_{\mathrm{LN}} = \Pi A_{\mathrm{LN}} \Pi^{-1}$ and $\widetilde{c}_{\mathrm{LN}} = \Pi c_{\mathrm{LN}}$. The scaling component is restricted to identity across LayerNorm boundaries to preserve normalization statistics.

\noindent\textbf{Residual stream and endpoints.}
We maintain consistent residual-stream permutations across all branches that are merged by residual addition, and fix the patch embedding input and classification head output with identity transforms ($Q_{\text{first}}=I$, $P_{\text{last}}=I$). Under these constraints, network-wise strict equivalence (Equation~\ref{eq:kcr-network-prop}) carries over from the CNN case. QC-LDPC coded quantization and ARI apply directly to the reparameterized transformer parameters without architecture-specific changes.

\subsubsection{Experimental Setup for ViT}
We evaluate \tool\ on two Vision Transformer models: DeiT-Tiny and DeiT-Small~\cite{touvron2021training}. 
DeiT-Tiny has about 5M parameters with 12 transformer blocks, 3 heads, and embedding dimension 192, while DeiT-Small has about 22M parameters with 12 blocks, 6 heads, and embedding dimension 384. 
We use official ImageNet-1K pretrained checkpoints and evaluate on ImageNet-1K directly. 
We also fine-tune both models on CIFAR-100 for 50 epochs using AdamW ($\mathrm{lr}=10^{-4}$) and a cosine schedule with 5-epoch warmup. 
CIFAR-100 images are resized to $224\times224$, and positional embeddings are interpolated following the standard DeiT fine-tuning protocol.

\noindent\textbf{Evaluation scope.}
This Transformer experiment is intended to validate the architectural compatibility and scaling behavior of \tool, rather than to repeat the full baseline comparison in Section~\ref{sec:4.4}. 
We therefore compare Base and \tool\ on DeiT models and do not include Aegis, BIN, or RA-BNN, since adapting these baselines to DeiT would require non-trivial retraining, architectural modification, or quantization-specific training.

\noindent\textbf{Attacks.}
We evaluate the three representative parameter-attack categories under a fixed budget for each attack.
For ProFlip~\cite{chen2021proflip}, we use a sparse flip budget of $B_f=100$.
For P3A~\cite{yao2020deephammer}, we perturb hidden dimensions in attention projections and MLP intermediate representations; to keep the structured attack setting comparable across DeiT-Tiny and DeiT-Small, we perturb 7 hidden dimensions for DeiT-Tiny and 14 hidden dimensions for DeiT-Small.
For APA~\cite{rakin2019bit}, we use $L_\infty=16/255$.
We do not repeat the joint adaptive attack here because joint-module adaptivity is already evaluated in Section~\ref{sec:joint-attack}; this ViT experiment focuses on Transformer compatibility and overhead scaling rather than a full attack-budget sweep.

\noindent\textbf{Metrics and configuration.}
We report clean ACC, ASR, model size, and latency (P50/P90).
\tool\ hyperparameters are unchanged from the CNN experiments for both DeiT-Tiny and DeiT-Small: 8-bit quantization, QC-LDPC rate $0.875$ with block size $N_b=128$, $\sigma_w=10^{-4}$, $M_s=5$, and $M_\ell=25$.
The confidence threshold $\tau$ is set to the 10th percentile of the clean confidence-margin distribution for each model.
Results are averaged over three seeds.

\begin{table*}[t!]
\setlength{\tabcolsep}{7.2pt}
\renewcommand{\arraystretch}{1.0}
\footnotesize
\centering
\caption{Transformer Generalization Evaluation on DeiT Models.}
\label{tab:vit-eval}
\begin{tabular}{c|c|c|c|ccc|c|c}
\hline\hline
\multirow{2}{*}{\textbf{Dataset}} 
& \multirow{2}{*}{\textbf{Model}} 
& \multirow{2}{*}{\textbf{Method}} 
& \multirow{2}{*}{\textbf{ACC (\%)}} 
& \multicolumn{3}{c|}{\textbf{ASR (\%)}} 
& \multirow{2}{*}{\textbf{\begin{tabular}{c}Model\\Size (MB)\end{tabular}}} 
& \multirow{2}{*}{\textbf{\begin{tabular}{c}Latency\\P50/P90\end{tabular}}} \\
\cline{5-7}
& & & & \textbf{ProFlip} & \textbf{P3A} & \textbf{APA} & & \\
\hline
\multirow{4}{*}{ImageNet-1K}
& \multirow{2}{*}{DeiT-Tiny}
& Base  & 72.20 & 96.74 & 91.38 & 57.63 & 22.83 & 7.45/9.82 \\
& 
& \tool & 70.83 & 43.52 & 54.17 & 18.94 & 6.72 & 7.91/10.64 \\
\cline{2-9}
& \multirow{2}{*}{DeiT-Small}
& Base  & 79.90 & 92.46 & 88.72 & 52.58 & 88.31 & 18.62/24.75 \\
& 
& \tool & 77.84 & 39.76 & 50.83 & 16.27 & 25.96 & 19.85/26.91 \\
\hline
\multirow{4}{*}{CIFAR-100}
& \multirow{2}{*}{DeiT-Tiny}
& Base  & 84.12 & 91.54 & 87.90 & 43.76 & 21.20 & 7.38/9.70 \\
& 
& \tool & 82.46 & 38.91 & 49.62 & 12.84 & 6.24 & 7.82/10.48 \\
\cline{2-9}
& \multirow{2}{*}{DeiT-Small}
& Base  & 88.37 & 89.92 & 85.35 & 40.28 & 82.85 & 18.35/24.30 \\
& 
& \tool & 85.91 & 36.48 & 47.29 & 11.36 & 24.36 & 19.54/26.38 \\
\hline\hline
\end{tabular}
\vspace{-3mm}
\end{table*}

\subsubsection{Results on ViT}

Table~\ref{tab:vit-eval} shows that \tool\ extends to the evaluated ViT-style Transformer models.
Across DeiT-Tiny and DeiT-Small on ImageNet-1K and CIFAR-100, \tool\ incurs only a 1.37--2.46 percentage-point drop in clean accuracy, indicating that the proposed defense preserves model utility on ViT-style architectures. 
At the same time, \tool\ consistently reduces ASR under all three parameter attack types. 
Specifically, ASR decreases from 89.92--96.74\% to 36.48--43.52\% for ProFlip, from 85.35--91.38\% to 47.29--54.17\% for P3A, and from 40.28--57.63\% to 11.36--18.94\% for APA.
These results suggest that KCR, QC-LDPC, and ARI remain effective against sparse, structured, and continuous parameter perturbations in the evaluated ViT-style architectures.

The deployment overhead also remains moderate. 
On ImageNet-1K, \tool\ reduces DeiT-Tiny from 22.83MB to 6.72MB and DeiT-Small from 88.31MB to 25.96MB. 
On CIFAR-100, where the fine-tuned checkpoints are smaller due to the reduced classification head, \tool\ reduces DeiT-Tiny from 21.20MB to 6.24MB and DeiT-Small from 82.85MB to 24.36MB. 
Latency increases only modestly across both datasets. 
We further discuss overhead scaling across CNNs and ViT-style models in Section~\ref{sec:transformer-overhead}.
}

\bin{
\subsection{Overhead Scaling}
\label{sec:transformer-overhead}

The Transformer evaluation further clarifies how \tool's overhead scales beyond CNN-based classifiers. 
Across CNNs and Transformers, the scaling trend is consistent because KCR is a preparation-time reparameterization with no per-query operations, QC-LDPC decoding is performed once at model loading and scales linearly with the number of encoded parameter blocks, and ARI is the main runtime cost controlled by $M_s$, $M_\ell$, and the slow-path triggering rate. 
Thus, within the evaluated CNN and ViT-style model scales, larger models increase absolute latency but do not necessarily amplify relative overhead.
\tool\ shows stable storage reduction and modest latency overhead across both architecture families. 
For CNNs, it reduces ResNet32 from 1.94--1.98MB to 0.57--0.58MB and VGG16 from 58.92--60.48MB to 17.33--17.78MB. 
For ViT-style models, it reduces DeiT-Tiny from 22.83MB to 6.72MB on ImageNet-1K and from 21.20MB to about 6.24MB on CIFAR-100; DeiT-Small is reduced from 88.31MB to 25.96MB on ImageNet-1K and from 82.85MB to about 24.36MB on CIFAR-100. 
These results indicate a consistent storage reduction of about 70\%, while P50/P90 latency increases remain modest across the evaluated CNN and ViT-style model scales.
}

\bin{
\subsection{Deployment Tiers}
\label{sec:deployment-tiers}

Having described the three modules of \tool, we now clarify how their protection composes under different platform trust assumptions. 
Because \tool\ relies on protected key material only for KCR's secret-key-based obfuscation, the framework degrades when TEE support is unavailable on low-cost edge or legacy devices. 
Table~\ref{tab:tiers} summarizes four deployment tiers.
{Tier~1 (TEE)} stores KCR keys and executes key-dependent preparation inside a TEE~\cite{mckeen2013intel,cerdeira2020sok}, providing the strongest guarantees by isolating both key material and key-dependent operations as assumed in Section~\ref{sec:threat-model}. 
{Tier~2 (HSM)} stores or wraps KCR keys inside a hardware security module~(HSM) with standard key-wrapping protocols~\cite{rfc3394}. 
This tier protects KCR key material at rest and during provisioning, but it does not provide TEE-equivalent isolation for arbitrary model-preparation code. 
Therefore, Tier~2 is intended for deployments in which the adversary may access stored checkpoints or deployment artifacts, but cannot observe the authenticated preparation service or read its process memory while KCR keys are being used. 
If a privileged online adversary can observe key use or recover the KCR key material, KCR's secret-key-based obfuscation guarantee no longer applies.
Tier~3 (TPM-sealed) binds keys to secure-boot measurements maintained by a Trusted Platform Module (TPM)~\cite{arbaugh1997secureboot}, resisting offline attackers and supply-chain tampering but not a fully compromised runtime. 
Thus, KCR's obfuscation holds only against checkpoint-level attackers who cannot unseal or observe the KCR keys, and is weakened or removed against privileged online adversaries.
{Tier~4 (no secure storage)} forfeits KCR's secret-key-based obfuscation but retains QC-LDPC correction and ARI stabilization; the remaining modules still provide load-time error correction/detection and inference-time stabilization.
Compared with retraining-based defenses such as Aegis~\cite{wang2023aegis}, \tool\ occupies a different deployment trade-off point. 
Aegis avoids trusted hardware but requires full retraining, architectural modification via dynamic exits, and is primarily designed for BFAs. 
\tool\ at Tier~1 assumes trusted key protection but requires no retraining, preserves the model architecture, and covers sparse, continuous, and structured parameter attacks. 
At Tiers~2--3, \tool\ can operate with alternative key-protection mechanisms but with weaker KCR guarantees; at Tier~4, it degrades to QC-LDPC correction and ARI stabilization without KCR obfuscation.

\begin{table}[t]
\setlength{\tabcolsep}{8.8pt}
\renewcommand{\arraystretch}{1.0}
\footnotesize
\centering
\caption{Deployment Tiers of \tool.}
\label{tab:tiers}
\begin{tabular}{l|l|c|c|c}
\hline\hline
\textbf{Tier} & \textbf{Key Protection} & \textbf{KCR} & \textbf{QC-LDPC} & \textbf{ARI} \\
\hline
1 & TEE (SGX/TrustZone) & Strong & \checkmark & \checkmark \\
2 & HSM                 & Partial & \checkmark & \checkmark \\
3 & TPM-sealed          & Partial    & \checkmark & \checkmark \\
4 & None                & --         & \checkmark & \checkmark \\
\hline\hline
\end{tabular}
\vspace{-4mm}
\end{table}

}

\bin{
\subsection{Key Lifecycle Management}
\label{sec:keymgmt}
Deploying \tool\ requires standard key-management support. A KCR key bundle $\{(\Pi_l^{(\mathrm{in})},D_l^{(\mathrm{in})}, \Pi_l^{(\mathrm{out})},D_l^{(\mathrm{out})})\}$ 
is generated per deployment instance, or per {(model, device)} tuple when per-device isolation is needed, inside a protected environment such as a TEE, HSM, or trusted provisioning service using a cryptographically secure PRNG. The bundle stores per-layer permutations and scaling factors; for the model scales considered here, its size remains small relative to the model checkpoint, ranging from tens of KB for ResNet32 to sub-MB scale for the evaluated ViT models. Bundles are provisioned after platform authentication, e.g., TEE remote attestation~\cite{menetrey2022exploratory} followed by an authenticated encrypted channel, or standard AES key wrapping such as RFC~3394. Rotation uses KCR's reversibility: the protected environment recovers the canonical parameter basis with the old bundle, samples a new bundle, reapplies KCR, and re-encodes the model with QC-LDPC, without retraining. Revocation is handled by key-version tags bound to the model artifact and checked at load time; stale versions are refused. 

}

\subsection{Threat-Model Alignment}
We evaluate threat-model alignment under a fixed ProFlip attack with a flip budget of 75. In the pre-load setting, bit flips are injected into the serialized checkpoint before loading, followed by QC-LDPC decoding and integrity verification at initialization. In the post-load setting, the same number of flips are applied directly to in-memory parameters after loading, bypassing load-time decoding, while all other configurations remain identical. 
Table~\ref{tab:threat-align-all} shows that under pre-load tampering, \tool\ maintains moderate ASR and substantially higher ACC across datasets and models, whereas post-load tampering leads to consistently higher ASR and severe accuracy degradation. This gap indicates that \tool’s robustness mainly stems from its load-time protection pipeline, while post-load tampering falls outside QC-LDPC’s correction scope. 
\bin{Accordingly, \tool\ targets at-rest parameter tampering; ARI nonetheless operates on every inference and provides per-query stabilization against residual runtime perturbations. Stronger runtime coverage would require complementary hardware-level mechanisms that address threats outside the model layer, such as ECC DRAM for transient memory-cell bit flips and confidential-computing primitives such as Intel TDX~\cite{intel2023tdx} for memory-integrity protection of the inference VM. These layers are orthogonal to \tool's model-layer defenses and are selected according to the deployment platform.}

\begin{table}[t!]
\centering
\setlength{\tabcolsep}{8pt}
\footnotesize
\caption{Pre-load vs. Post-load Tampering.}
\label{tab:threat-align-all}
\renewcommand{\arraystretch}{1.0}
\begin{tabular}{c|c|cc|cc}
\hline\hline
\multirow{2}{*}{\textbf{Dataset}} &
\multirow{2}{*}{\textbf{Model}} &
\multicolumn{2}{c|}{\textbf{Pre-load}} &
\multicolumn{2}{c}{\textbf{Post-load}} \\
\cline{3-6}
 &  & \textbf{ASR} & \textbf{ACC} & \textbf{ASR} & \textbf{ACC} \\
\hline
\multirow{2}{*}{CIFAR-10}
 & ResNet32 & 45.02 & 50.16 & 70.36 & 27.35 \\
 & VGG16    & 43.97 & 51.27 & 68.16 & 28.39 \\
\hline
\multirow{2}{*}{CIFAR-100}
 & ResNet32 & 47.36 & 32.69 & 72.93 & 17.17 \\
 & VGG16    & 44.25 & 37.40 & 69.24 & 20.63 \\
\hline
\multirow{2}{*}{Tiny-ImageNet}
 & ResNet32 & 48.47 & 26.52 & 73.42 & 13.65 \\
 & VGG16    & 49.19 & 29.17 & 74.88 & 14.82 \\
\hline\hline
\end{tabular}
\vspace{-4mm}
\end{table}

\subsection{Threats to Validity}

\noindent\textit{Internal validity.}  
Our evaluation uses specific implementations of ProFlip, P3A, and APA with adaptive variants. Although strong settings from prior work (including EOT and projection-aware optimization) are adopted, attack strength may vary with alternative heuristics or parameter choices. In addition, fixed design choices (e.g., LDPC rate, block size, KCR initialization, and ARI sampling) may not cover the full configuration space, potentially affecting absolute robustness while preserving comparative trends.

\noindent\textit{External validity.}  
\bin{The main evaluation is conducted on CIFAR-10/100 and Tiny-ImageNet using ResNet32 and VGG16, and we further include DeiT-Tiny/Small on ImageNet-1K and CIFAR-100 to examine Transformer compatibility and overhead scaling. However, results may still differ for substantially larger models, generative models, LLMs, and deployment-specific factors such as key management, runtime integrity mechanisms, and hardware-level faults. Extending \tool\ to LLM-scale and generative-model deployments, with token-level serving latency, generation-quality metrics, and model-specific attack surfaces, remains important future work.}

\subsection{Limitations}
\tool\ combines parameter-space transformations and redundancy to improve robustness against sparse, continuous, and structured parameter attacks while maintaining high accuracy and moderate deployment overhead. Practical deployment introduces challenges including secure key management, residual QC-LDPC decoding errors, and additional latency from inference redundancy.
\bin{\tool's strongest guarantees require a TEE at Tier~1 (Section~\ref{sec:deployment-tiers}); at Tiers~2--3, alternative primitives (HSM, TPM-sealed storage) provide weaker key-protection guarantees for KCR, while at Tier~4, \tool\ retains QC-LDPC correction and ARI stabilization but forfeits KCR obfuscation. Post-load runtime tampering falls outside the load-time verification scope of QC-LDPC; ARI provides partial per-inference mitigation, but full coverage of runtime memory threats belongs to the hardware layer (ECC DRAM, Intel TDX) rather than the model layer addressed by \tool. Our evaluation primarily focuses on clean accuracy and attack success rate, which directly measure utility preservation and parameter-attack robustness. Although ARI uses confidence margins for slow-path escalation, it is designed to stabilize predictions under parameter perturbations rather than to calibrate probabilistic outputs. We leave evaluation with ECE, NLL, and shifted input distributions to future work.}
In addition, adversaries may induce decoding failures or trigger redundancy to increase overhead and affect availability, which requires deployment-level mitigation.
Although we evaluate representative sparse, continuous, and structured attacks, highly adaptive strategies that target defense components or approximate transformation effects may pose additional challenges. Finally, compatibility with diverse inference stacks, deployment pipelines, and large-scale key management remains an important direction for future work.

\section{Conclusion}

We introduced \tool, a generalized defense for deep neural networks against parameter attacks. By combining reparameterization, coded quantization, and adaptive inference, \tool\ delivers robust protection across sparse, continuous, and structured attacks while maintaining high model performance and incurring minimal deployment overhead without retraining. These results suggest that \tool\ is a practical defense option for securing DNN deployments against at-rest parameter tampering.

\bibliographystyle{IEEEtran}
\bibliography{acm}

\begin{IEEEbiography}[{\includegraphics[width=1.25in,height=1.25in,clip,keepaspectratio]{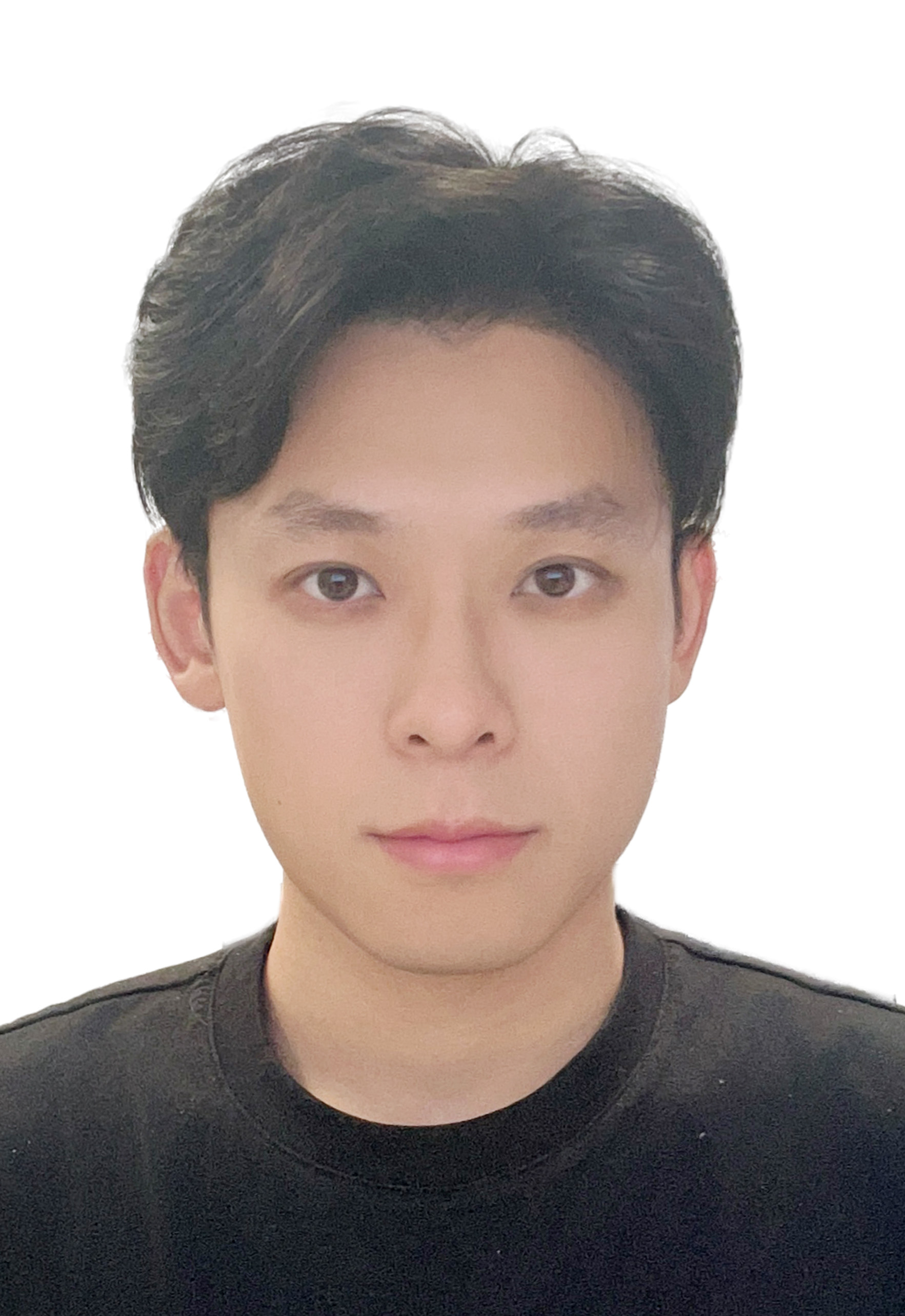}}]{Bin Duan} received the master's degree from Southeast University, China. He is currently pursuing the Ph.D. degree with the School of Electrical Engineering and Computer Science, The University of Queensland, Australia. His research interests include AI security, software testing for AI systems, and dependable AI.
\end{IEEEbiography}
\vspace{-13cm}
\begin{IEEEbiography}
[{\includegraphics[width=1.25in,height=1.25in,clip,keepaspectratio]{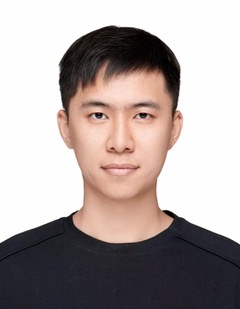}}]{Zeyu Bai} received the master's degree in Computer Science from The University of Queensland (UQ), Australia. He is currently working as a Platform Operations Engineer. His research and professional interests include neural network security, information system management, and system operation management.
\end{IEEEbiography}
\vspace{-13cm}
\begin{IEEEbiography}
[{\includegraphics[width=1.25in,height=1.25in,clip,keepaspectratio]{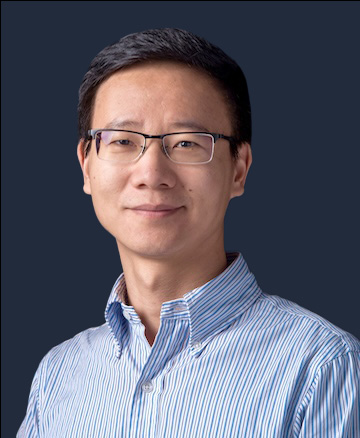}}]{Guowei Yang} received his PhD in electrical and computer engineering from the University of Texas at Austin, USA. He is a Senior Lecturer in Software Engineering in the School of Electrical Engineering and Computer Science at the University of Queensland. His research interests lie in software engineering, and its synergy with artificial intelligence and programming languages, with a focus on improving reliability and security of both traditional software and AI systems. His work has been published in top-tier conferences and prestigious journals such as ICSE, FSE, ASE, PLDI, TSE, and TOSEM. He actively contributes to the academic community as a program committee or organizing committee member for esteemed conferences such as ICSE, FSE, ASE, ISSTA, and DSN.
\end{IEEEbiography}

\end{document}